%% file: qh_idmrg.tex
\documentclass[prl,reprint,twocolumn,times,showpacs,superscriptaddress]{revtex4-1}

\usepackage{amsmath, amssymb, braket, dsfont}
\usepackage[mathscr]{euscript}
\usepackage{graphicx}
\usepackage{subfigure}		
\usepackage{epstopdf}
\usepackage[matrix,arrow,curve]{xy}		
\usepackage[breaklinks]{hyperref}

\newcommand{\gsd}{{\mathfrak{m}}}		
\newcommand{\dxp}[1]{{\langle\!\langle{#1}\rangle\!\rangle}}	
\newcommand{\bondop}[1]{{\mkern 3.5mu\overline{\mkern-3.5mu{#1}\mkern-1.1mu}\mkern 1.1mu}}
\newcommand{\drehen}{\bondop{G}}				
\newcommand{\HallVis}{{\eta_{_H}}}			
\newcommand{\shift}{\mathscr{S}}

\usepackage[usenames,dvipsnames]{color}			
\definecolor{purple}{rgb}{0.5,0,0.5}

\begin{document}

\title{Topological characterization of fractional quantum Hall ground states from microscopic Hamiltonians}
\author{Michael P. Zaletel}
\affiliation{Department of Physics, University of California, Berkeley, California 94720, USA}
\author{Roger S. K. Mong}
\affiliation{Department of Physics, University of California, Berkeley, California 94720, USA}
\affiliation{Department of Physics, California Institute of Technology, Pasadena, California 91125, USA}
\author{Frank Pollmann}
\affiliation{Max-Planck-Institut f\"ur Physik komplexer Systeme, 01187 Dresden, Germany}

\begin{abstract}

We show how to numerically calculate several quantities that characterize topological order starting from a microscopic fractional quantum Hall (FQH) Hamiltonian.
To find the set of degenerate ground states, we employ the infinite density matrix renormalization group (iDMRG) method based on the matrix-product state (MPS) representation of FQH states on an infinite cylinder.
To study localized quasiparticles of a chosen topological charge, we use pairs of degenerate ground states as boundary conditions for the iDMRG.
We then show that the wave function obtained on the infinite cylinder geometry can be adapted to a torus of arbitrary modular parameter, which allows us to explicitly calculate the non-Abelian Berry connection associated with the modular $\mathcal{T}$-transformation.
As a result, the quantum dimensions, topological spins, quasiparticle charges, chiral central charge, and Hall viscosity of the phase can be obtained using data contained entirely in the entanglement spectrum of an infinite cylinder.

\end{abstract}

\maketitle
\phantomsection
\addcontentsline{toc}{section}{Main text}


Over the last decades several new kinds of phases have been discovered that cannot be characterized by spontaneous symmetry breaking but instead exhibit \emph{topological order} \cite{Wen-1990,WenNiu-1990}. 
A prominent example of a topological ordered phase is the fractional quantum Hall (FQH) effect \cite{Tsui-1982}.
These systems support quasiparticles (QPs) with exotic exchange statistics (i.e., they are neither fermions nor bosons)
	and have been proposed as a platform for a ``topological'' quantum computer \cite{Kitaev-1997, DasSarmaInterferometer-2005, BondersonInterferometer-2006, Nayak-2008}.
Exact diagonalization (ED) numerics have played a decisive role in the study of FQH phases \cite{Laughlin-1983,Yoshioka-1983,Haldane-1983},
	and many characteristics of the topological order can be extracted directly from the ground states (GSs) via their entanglement structure \cite{KitaevPreskill, Levin-2006, Li-2008, Laeuchli-2010, Papic-2011, Zhang-2012}.
However, the exponential growth of the Hilbert space as a function of particle number means that ED is prohibitively expensive beyond \mbox{$\approx\hspace{-0.4ex}20$} particles.
While it is possible to obtain very accurate results for some systems with a small correlation length, such as the $\nu=1/3$ Coulomb state,
	more complicated systems (system with higher Landau level, hierarchy states, or non-Abelian phases) are much harder to access and finite size effects are much stronger.

\begin{figure}[t]
	\includegraphics[width=8cm]{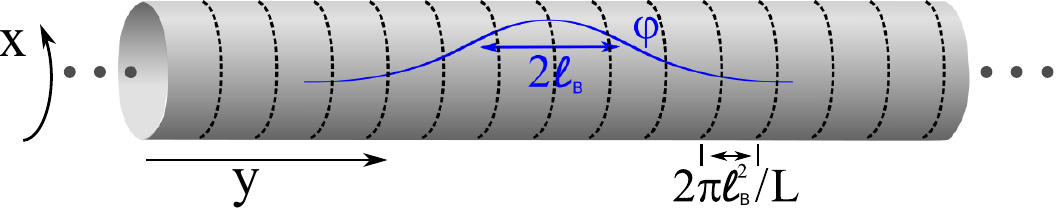}
	\caption{Single particle states $\varphi(x,y)$ in the lowest Landau level of an infinite cylinder.
		$L$ is the circumference of the cylinder and $\ell_B$ is the magnetic length.				}
	\label{fig:cylinder}
\end{figure}

In this paper we address the limitations of ED by using the infinite density matrix renormalization group (iDMRG) algorithm to obtain the matrix product state (MPS) representation of the GSs of FQH Hamiltonians on \emph{infinitely} long cylinders with finite circumference $L$ (see Fig.~\ref{fig:cylinder}). 
The space of MPSs has been shown to be an exact and efficient representation of the model QH states \cite{Zaletel-2012,estienne-2012}; iDMRG extends these results to non-model states by variationally optimizing an MPS with respect to a microscopic Hamiltonian \cite{White-1992}.
For a system of size $L_x \times L_y$, \emph{finite} DMRG reduces the computational complexity from $\mathcal{O}(b^{L_x L_y})$ via ED to $\mathcal{O}(L_x L_y b^{L_x})$, with $b\gtrsim1$; taking $L_y \to \infty$ using \emph{infinite} DMRG gives $\mathcal{O}(b^{L_x})$.
Several groups have by now implemented finite DMRG to simulate FQH Hamiltonians \cite{Naokazu-2001,BergholtzKarlhede-2003,Feiguin-2008,Kovrizhin-2010,Zhao-2011, Hu-2012}.
The infinite cylinder geometry has several additional numerical advantages, such as translation invariance both along and around the cylinder, zero curvature effects \cite{Hu-2012}, and the absence of gapless edge excitations that slow down the convergence.
In contrast to the standard bipartition of the torus \cite{LauchliBergholtz-2010,Laeuchli-2010,ZhaoBergholtz2012}, cutting the infinite cylinder gives the entanglement spectrum of a single edge.

We demonstrate how to obtain various characterizing quantities of quantum Hall systems from microscopic Hamiltonians using the iDMRG simulations.
The most important step is to use iDMRG to systematically obtain the full set of degenerate ground states on an infinite cylinder, starting from a Hamiltonian.
Their entanglement spectra encode the quantum dimensions $d_a$ \cite{KitaevPreskill,Levin-2006,Dong-2008} and charges $Q_a$ of the quasiparticles.
Topological quasiparticles on a cylinder appear as domain walls between pairs of degenerate ground-states, which we numerically optimize in order to obtain the energy and MPS of a localized quasiparticle of a chosen topological charge.
We apply the infinite cylinder technique to spin-polarized electrons at filling $\nu = 1/3$, $2/5$ and $1/2$ for a range of different Haldane pseudopotentials.

	Several topological properties of the FQH state manifest themselves on a torus rather than on an infinite cylinder.
To construct the MPS for the ground-states of a torus with a twist in it, we cut out a segment of the infinite cylinder MPS and insert a ``twist operator'' before gluing the ends back together to form a ring, which is justified so long as the correlation length is small compared to the new circumference.
By adiabatically changing the twist we can compute the non-Abelian Berry phase $U_T$ associated with a modular $\mathcal{T}$-transformation, which determines the topological spins $\theta_a$ of the quasiparticles, the central charge $c$ of the edge theory, and the Hall viscosity $\HallVis$ of the bulk \cite{Avron-QHViscosity95,Read-QHViscosity09}.
Remarkably the result again depends only on the entanglement spectrum of the infinite cylinder.

Recently Cincio and Vidal \cite{Cincio-2012} applied iDMRG to a lattice Hamiltonian on a cylinder and extracted the braiding statistics of the anyons (``$\mathcal{S}$'' and ``$\mathcal{T}$'') using wavefunction overlaps \cite{Zhang-2012}.
In this work, we employ iDMRG to continuum FQH Hamiltonians and extract topological information purely from the entanglement spectra.

\paragraph{Model and method.} 
We consider the cylinder geometry \cite{Rezayi-1994,Bergholtz-2005,Seidel-2005} with a coordinate $x$ running around the circumference of length $L$, and $y$ running along the infinite length of the cylinder (see Fig.~\ref{fig:cylinder}).
The Landau gauge $\mathbf{A} = \ell_B^{-2}(-y,0)$ conserves the $x$-momentum around the cylinder.
The orbitals in the first Landau level are
\begin{align}
	\varphi_n(x, y) = \frac{e^{ i k_n x - \frac{1}{2 \ell_B^2} (y - k_n \ell_B^2)^2}}{\sqrt{L \ell_B \pi^{1/2}}}, \quad k_n = \frac{2 \pi n}{L},
\end{align}	
where $n \in \mathbb{Z}$, with a density of one orbital per flux quanta.
Because each orbital is localized at $y_n=k_n \ell_B^2$, we treat the system as a one-dimensional chain.
The most general two-body interaction allowed by translational symmetry can be specified by coefficients $V_{km}$,
\begin{align}
\label{h}
	\hat{H} = \sum_n \sum_{k \geq |m|} V_{km} c^{\dag}_{n + m} c^\dagger_{n+k} c^{\vphantom{\dag}}_{n+m+k} c^{\vphantom{\dag}}_n.
\end{align}
If the interactions are local in real space, they will be Gaussian localized with a range on the order of $L/\ell_B$ in the chain representation.
The Hamiltonian can be parametrized by the set of ``Haldane pseudopotentials'' $V_m$ which can represent any rotationally and translationally invariant two-body interaction within a single Landau level. For example, the hard core Haldane pseudopotential $V_1$ takes the form $V_{km} \propto (k^2 - m^2)e^{-\frac{1}{2}(2 \pi \ell_B / L)^2(k^2+m^2)}$ \cite{Rezayi-1994}.
To represent $\hat{H}$ at some fixed accuracy we must keep $\mathcal{O}(L^2/\ell_B^2)$ terms in the Hamiltonian.

For our numerical simulations, we use the iDMRG algorithm \cite{McCulloch-2008} which is based on the infinite MPS representation,
\begin{align}
	\ket{\Psi} &= \!\! \sum_{ \{ j_n \} } \!\!
		\left[ \cdots B^{[0]j_0} B^{[1]j_1} \cdots \right]
		\ket{\ldots, j_0, j_1, \ldots}
	,  \label{eq:mps}
\end{align}
where $B^{[n]j_n}$ are $\chi\times\chi$ matrices and $\ket{j_n}$, $j_n\in\{0,1\}$ represent the occupancy at orbital $n$.
Assuming the state $\ket{\Psi}$ is translationally invariant with a unit cell of length $M$, then we need only store $M$ different tensors $B^{[i]}$ to express the MPS, \textit{i.e.},  $B^{[i]}=B^{[i + M]}$; at filling $\nu=p/q$, $M$ must be a multiple of $q$.
MPSs have proven to be extremely successful in the simulation of gapped, one-dimensional systems because their GSs can be expressed to very high accuracy by keeping a relatively small $\chi$ even for infinite system size \cite{Hastings-2007, Gottesman-2009, Schuch-2008, Nakamura-2012}.
The iDMRG algorithm proceeds by iteratively minimizing $E = \braket{\Psi | \hat{H} | \Psi}$ within the space of MPS.
The $\chi$ value needed to express the ground state to a given accuracy grows exponentially with $L$; a moderate bond dimension of $\chi=3600$ was sufficient for the largest system considered here, which took under a day.
In the Supplementary material we address several technical issues particular to QH iDMRG
	\footnote{In the Supplemental materials we explain the numerical issues particular to quantum Hall iDMRG, including the MPO construction and ergodicity issues.
		We also detail the computation of the quasiparticle charges, flux matrices, and modular $\mathcal{T}$-matrix from the entanglement spectrum.
		}.


	After each variational optimization, the wavefunction is projected back to the original MPS form by keeping only the most important contributions in the Schmidt decomposition of each bond $\bar{n}$,  
\begin{equation}
	\ket{\Psi} = \sum_{\alpha} \lambda_{\bar{n}; \alpha} \ket{\alpha}_L \ket{\alpha}_R.
\end{equation}
The Schmidt states $\ket{\alpha}_{L/R}$ form orthonormal bases for the sites to the left and right of the bond $\bar{n}$, and the Schmidt values are $\lambda_{\bar{n}; \alpha}$.
The entanglement entropy for this bipartition is directly obtained from the Schmidt values $S_{\bar{n}} = -\sum_{\alpha} \lambda_{\bar{n}; \alpha}^2 \log \lambda_{\bar{n}; \alpha}^2 $.

We incorporate both particle number and momentum conservation of $\ket{\Psi}$ in its MPS representation and DMRG algorithm \cite{Schollwock2011}, which assume an important role in the MPS's subsequent analysis. 
We incorporate both particle number and momentum conservation in the MPS representation and iDMRG algorithm \cite{Schollwock2011}, which assume an important role in the subsequent analysis. 
The quantum numbers are defined to be
\begin{subequations}\label{eq:charges}\begin{align}
	\hat{C} &= \sum_n \hat{C}_n \equiv \sum_{n}  (\hat{N}_n - \nu ) \quad \textrm{(particle number)}	, \\
	\hat{K} &= \sum_n \hat{K}_n \equiv \sum_{n} n (  \hat{N}_n - \nu ) \quad \textrm{(momentum)}	,
\end{align}\end{subequations}
where $\hat{N}_n$ is the number operator at site $n$.

The Schmidt states $\ket{\alpha}_{L/R}$ of $\ket{\Psi}$ on bond $\bar{n}$ form orthonormal bases for the sites to the left and right of the bond, with corresponding Schmidt values $\lambda_{\bar{n}; \alpha}$.
They have a well-defined particle number $\bondop{C}_{\bar{n}; \alpha}$ representing the total charge to the left of bond $\bar{n}$.
We will view $\bar{C}_{\bar{n}}$ as a diagonal matrix acting in the set of Schmidt states (as for $\bar{K}$).
It will prove useful to define a ``bond expectation value,'' $\braket{\bondop{C}_{\bar{n}}} \equiv \sum_{\alpha} \lambda^2_{\bar{n}; \alpha} \bondop{C}_{\bar{n}; \alpha}$	,
which gives the expected value of the charge to the left of bond $\bar{n}$ (as for $\bondop{K}$).
The Schmidt values and their quantum numbers $(\lambda, \bar{C}, \bar{K})_{\bar{n}}$ constitute the ``orbital entanglement spectrum'' (OES) of the bond.
The corresponding entanglement entropy is defined as $S_{\bar{n}} \equiv -\sum_{\alpha} \lambda_{\bar{n}; \alpha}^2\log \lambda_{\bar{n}; \alpha}^2$. 

\begin{figure}[t]
	\includegraphics[width=8.5cm]{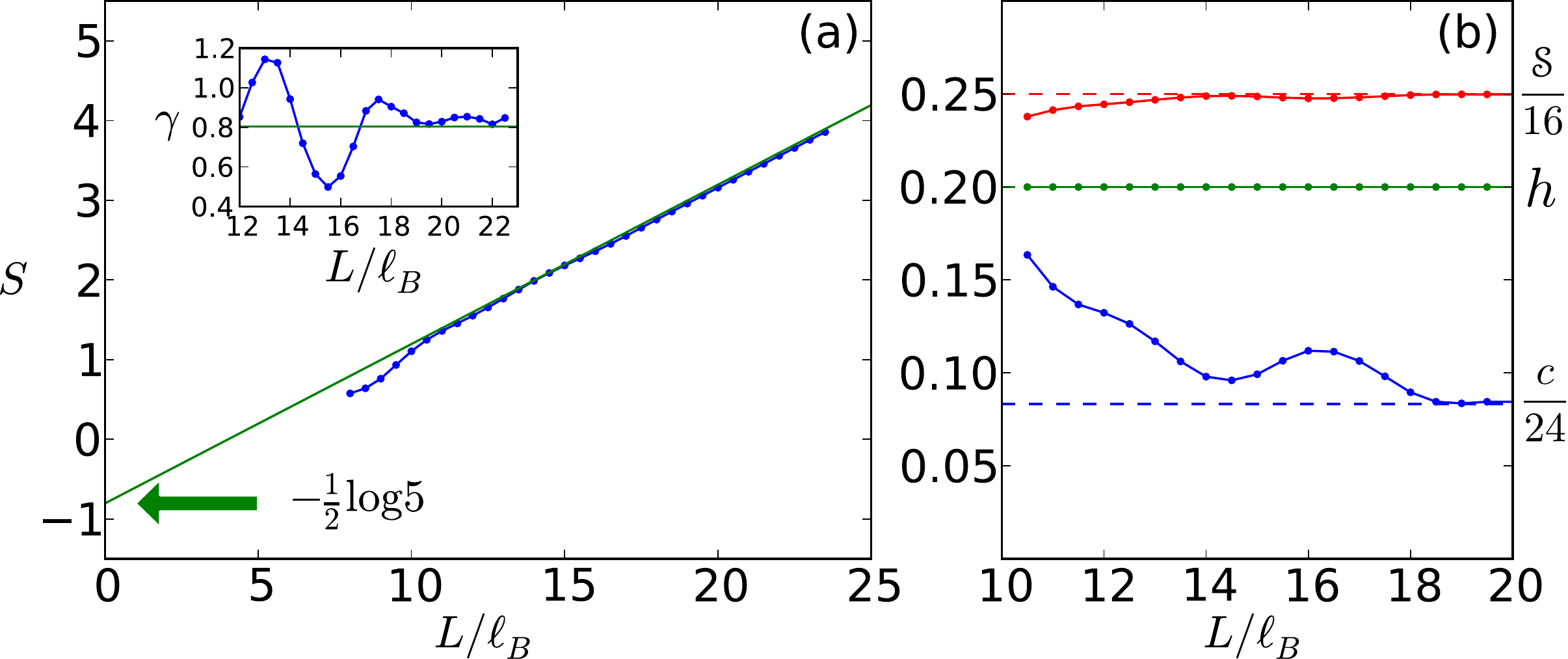}
	\caption{%
		(a) Entanglement entropy $S$ of the $\nu=2/5$ state as a function of the circumference $L$, for a bipartition into two half-infinite cylinders.
		The inset shows the estimated TEE $\gamma = L\frac{dS}{dL} - S(L)$ converging to $\gamma = \frac12 \log 5$ (green line) for large $L$.
		(b) Estimate of the chiral central charge $c/24$, the topological spin $h$ of the $e/5$ quasiparticle, and the Hall viscosity as expressed through the `shift,' $\shift = \frac{8\pi \ell_B^2}{\hbar \nu} \HallVis$ \cite{Read-QHViscosity09}.
		Dashed lines show expected hierarchy values of $2/24, 1/5, 4/16$ respectively.
		That $h$ is identically correct is peculiar to Abelian quasiparticles.
	\label{fig:25topo} 
	}

\end{figure}

\paragraph{Topological order and quasiparticles.}
We now recall some basic facts about topological order on a cylinder \cite{MooreRead1991}.
A topologically ordered state with $\gsd$ quasiparticle types (labeled by ``$a$'') has an $\gsd$-fold GS degeneracy on an infinite cylinder.
The chiral CFT describing the edge contains $\gsd$ scaling operators $\{\phi_a\}$, which insert a corresponding QP $a$ near the edge.
The Hilbert space of the edge $\mathcal{H}$ can be decomposed into a direct sum of subspaces $\mathcal{H}_a$ that contain the edge states with topological charge $a$.
In addition,  the states of the low-lying OES are in one-to-one correspondence with the states of the edge CFT \cite{KitaevPreskill,Li-2008}.
With an entanglement cut running around the circumference of the cylinder, we can choose a particular basis $\{ \ket{\Xi_a} \}$ for the $\gsd$-dimensional vector space of GSs such that the low-lying part of the OES of $\ket{\Xi_a}$ contains states \emph{only} in $\mathcal{H}_a$.
As each $\ket{\Xi_a}$ contains fewer states in the low-energy part of the OES than in the full CFT, they have lower entanglement entropy and are referred to as the `minimal entanglement states' (MESs) \cite{Zhang-2012}.
Translations permute the MESs, so the OES of a state depends on the bond $\bar{n}$ between adjacent orbitals where the cylinder is cut. Consequently each bond $\bar{n}$ is labeled by the sector $a$ found in its OES; we therefore denote the corresponding bond by $\bar{a}$
	\footnote{We note that for bosonic states the $\mathds{1}$ sector will not appear in the OES, a subtlety we address in the Supplemental material.
For the fermionic states studied here, $\mathds{1}$ will.}%
	.

	We now outline the procedure for finding the full set of degenerate GSs using iDMRG.
On an infinite cylinder the MESs are energy eigenstates, possibly with a degeneracy split exponentially in $L$ \cite{Cincio-2012}. 
For QH problems in a topological phase, we expect initializing the iDMRG using different orbital configurations ``$\mu$'' (for example, $\mu = 010$ gives $\ket{\cdots010010\cdots}$) should generate the set of distinct MES after iDMRG optimization.
For many QH  states, this is a consequence of MESs' distinct quantum numbers $K$, though we suspect the result is more general (Supplemental material \cite{Note1}).
If we find optimized energies such that $E_{\mu_1} < E_{\mu_2}$, then the state derived from initial state $\mu_2$ is rejected. The iDMRG is the numerical check that a given $\mu$ leads to one of the $\gsd$ GSs.
The orbital string $\lambda_a$ associated to the MES $\ket{\Xi_a}$ is called the state's `root configuration' or  `pattern of zeros' \cite{BernevigHaldane2008, WenWang-2008}.

	The entanglement entropy of each MES $\ket{\Xi_a}$ scales with the circumference as $S \approx sL - \gamma_a$, where $\gamma_a=\log \sqrt{\sum_{b} d_{b}^2} - \log(d_a)$ are the topological entanglement entropies (TEEs) and $d_{a}$ are the quantum dimensions of the QPs \cite{KitaevPreskill, Levin-2006, Dong-2008}.
From the TEEs $\gamma_a$ we can determine if we have the complete set of MESs \cite{Cincio-2012}.
As an example of an Abelian model ($d_a = 1$ for all $a$), we consider $\nu=2/5$ filling for Haldane pseudopotentials $V_3/V_1 = 0.05$, for which the seed $\ket{01010}$ and its translates were numerically determined to provide the five MESs.
This has been considered before in an ED study with Coulomb interactions on a torus for circumferences up to $L=18\ell_B$ \cite{Laeuchli-2010}, from which it was difficult to determine $\gamma$ accurately.
We are able to go up to $L=23.5\ell_B$ and obtain the results shown in Fig.~\ref{fig:25topo}.
The estimated TEE is $\gamma\approx 0.83$ (close to $\frac12 \log5 \approx 0.8047$).

\begin{figure}[t]
	\begin{minipage}{52mm}
		\includegraphics[width=\textwidth]{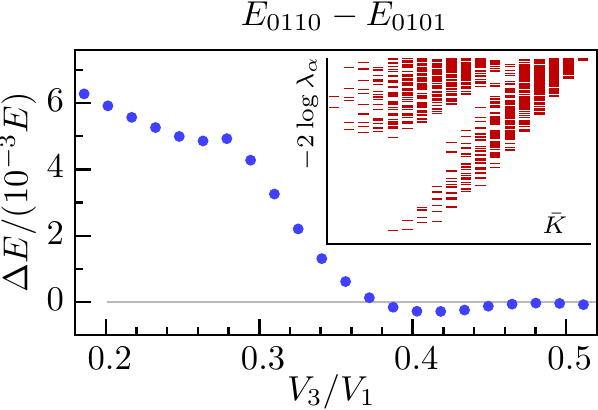}
		\\[-2ex]\hspace{-\textwidth}\begin{minipage}{0mm}\vspace{-72mm}\subfigure[]{\label{fig:esplit}}\hspace{-27mm}\end{minipage}
		\end{minipage}
	\,
	\begin{minipage}{31mm}
		\includegraphics[width=\textwidth]{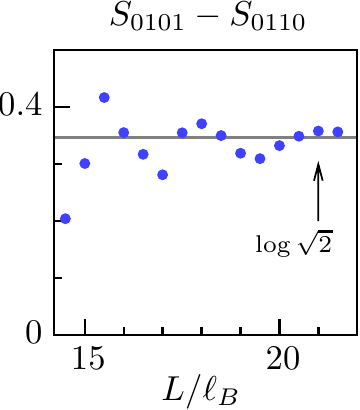}
		\\[-2ex]\hspace{-\textwidth}\begin{minipage}{0mm}\vspace{-72mm}\subfigure[]{\label{fig:ssplit}}\hspace{-27mm}\end{minipage}
		\end{minipage}
	\caption{%
		(a) The relative energy difference between states in the $0110$ and $0101$ sectors at $\nu=1/2$ filling and $L = 16 \ell_B$, plotted against the ratio of pseudopotential strengths $V_3/V_1$.
			At small $V_3/V_1$, the energies $E_{0110} > E_{0101}$ and hence there is only a twofold GS degeneracy (CFL phase), whereas at large $V_3/V_1$, the energies are roughly equal, which gives rise to a six fold GS degeneracy (MR phase).
			Inset shows the real-space entanglement spectrum of the $0101$ state plotted versus $\bondop{K}$, with counting 1, 2, 4, 8, 14.
		(b) Fixing a point $V_3/V_1 = 0.4$ in the MR phase, we increase $L$ and measure the difference in entanglement entropies $S_{0101} - S_{0110}$.
			The result is consistent with $d_\sigma = \sqrt{2}$ for the quantum dimension of the $e/4$ quasiparticle.
	}
\end{figure}		

As an example that has a non-Abelian phase, we consider the filling $\nu = 1/2$ that contains both the gapless composite Fermi liquid (CFL) phase and the gapped, non-Abelian MR phase \cite{MooreRead1991,HaldaneRezayi2000}.
Using a sum of Haldane pseudopotentials $V_1$ and $V_3$, we tune between the CFL (small $V_3/V_1$) and MR (intermediate $V_3/V_1$) phases.
We start the iDMRG either with the $\ket{0110}$ configuration (which provides four states via translation) or $\ket{0101}$ (which provides two).
We observe that in the suspected CFL phase the energies of the two sectors are split, whereas the MR phase is nearly sixfold degenerate, as illustrated in Fig.~\ref{fig:esplit}.
Fixing a point in the MR phase, the difference in the entanglement entropies of the MESs is $S_{0101} - S_{0110} \approx 0.36$, which implies that the $S_{0101}$ state is associated with a non-Abelian particle of quantum dimension $1.43$  (close to that of the $e/4$ excitation, $\sqrt{2}\approx 1.41$), as illustrated in Fig.~\ref{fig:ssplit}.
This supports the non-Abelian nature of the state.

\begin{figure}
	\includegraphics[width=8cm]{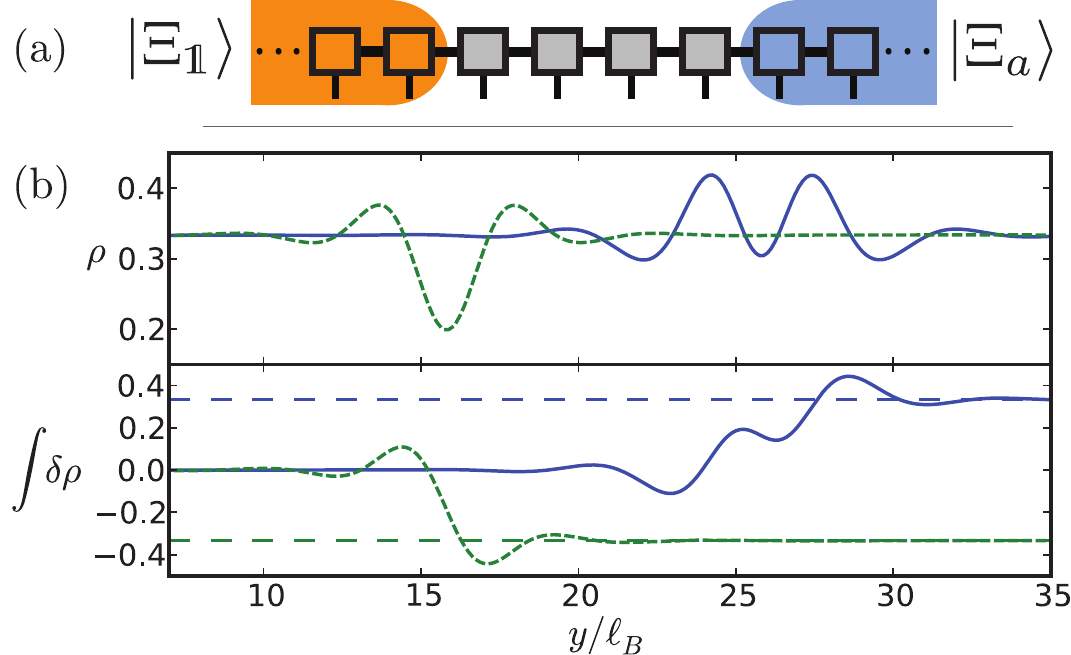}
	\caption{%
		(a) The MPS used to represent a quasiparticle `$a$'.
		(b) The charge density $\rho(y)$ of the numerically optimized QP with charge $+e/3$ (blue line)  and $-e/3$ (green line) of the $\nu = 1/3$ state.
			The cylinder has circumference $L = 16 \ell_B$, and the potential approximates a dipolar $r^{-3}$ interaction.
	}
	\label{fig:qp} 
\end{figure}		

In the cylinder geometry, the QPs appear as domain walls between the degenerate ground-states.
To generate a quasiparticle of type $a$ in the vicinity of $y = 0$, we use the $B$-matrices of the identity MPS $\ket{\Xi_{\mathds{1}}}$ for sites at $y \ll 0$ and the $B$-matrices of $\ket{\Xi_{a}}$ for $y \gg 0$.
In the vicinity of $y=0$, we insert a finite number of $B$-matrices which we numerically optimize.
For efficiency, we work with QPs of fixed momentum $K$; the resulting QPs for the $\nu = 1/3$ state are illustrated in Fig.~\ref{fig:qp}.
To calculate the charge of the particle we measure $Q_a=\lim_{\epsilon \to 0}\sum_{n}e^{- \epsilon |n| }\hat{C}_n$, where $\hat{C}_n$ is the on-site charge operator defined in Eq.~\eqref{eq:charges}.
The resulting charge is in fact \emph{independent} of the $B$-matrices used to glue together the MPS;
	$Q_a$ can be calculated exactly knowing only the OES of the infinite MPS $ \ket{\Xi_{a}}$:
\begin{align}
	\label{eq:qpcharge}
	e^{2 \pi i Q_a} &= e^{ 2 \pi i \left( \dxp{\bondop{C}} - \bondop{C}_{\bar{a}} \right)} ,
\end{align}
where $\dxp{\bondop{C}}$ is the average of $\braket{\bondop{C}_{\bar{n}}}$ over \emph{all} bonds $\bar{n}$ in the MPS of $\ket{\Xi_a}$ \cite{Note2}.
This expression is exact and reduces to a known result in the limit of a thin cylinder \cite{WenWang-2008,BernevigHaldane2008}.

\begin{figure}
	\includegraphics[width=8cm]{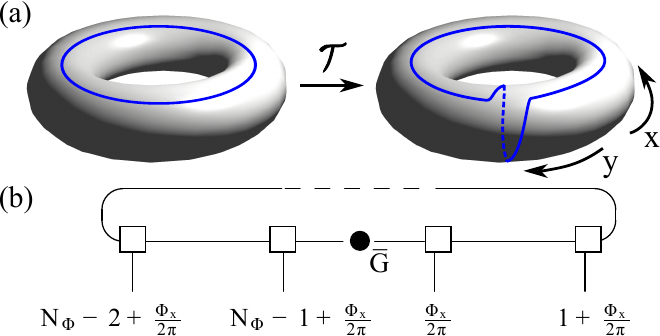}
	\caption{%
		(a) A torus with dimensions $L_x\times L_y$.
			Taking the modular parameter $\tau \to \tau+1$ generates a modular transform ``$\mathcal{T}$'' of the torus and acts on the set of GSs as a matrix $U_T$.
		(b) The wavefunction on the torus is represented by a periodic MPS.
			The sites are labeled by $n \in \mathbb{Z} + \frac{\Phi_x}{2\pi}$ due to the flux threading in the $x$-direction.
	}
	\label{fig:torus}
\end{figure}

\paragraph{Ground states on a torus and topological spin.}
Several quantities of interest, such as the phase's response to flux insertion and modular transformations, are defined on the torus geometry, so it is useful to convert the cylinder MPS to a torus MPS.
The torus is made by taking a cylinder of circumference $L_x$ and identifying points $(x,y)$ to $(x+\tau_x L_x, y+L_y)$, where $\tau = \tau_x+iL_y/L_x$ is the modular parameter.
We also allow fluxes $\Phi_{x/y}$ to thread through the two cycles.
For a suitable definition of the $N_\Phi = L_x L_y/2\pi\ell_B^2$ orbitals of the torus (Supplemental material \cite{Note1}), we can construct wavefunctions on the torus from those on the infinite cylinder by taking a finite segment of the cylinder MPS and connecting the two edge auxiliary bonds together to form a ring \cite{Cincio-2012}, as illustrated in Fig.~\ref{fig:torus}(b).
There is no need to reoptimize the $B$-matrices near the seam: locally, the torus Hamiltonian is identical to that of an infinite cylinder, and if $L_y$ is greater than the correlation length the periodic MPS will have the same local correlations as the iMPS.
The fluxes and modular parameter can be accounted for by inserting a diagonal matrix $\drehen$ when connecting the two edge auxiliary bonds ( Supplemental material\cite{Note1}),
\begin{align}
	\label{eq:drehen_main}
	\drehen &= (-1)^{(N^e-1)\bondop{C}} \, e^{-2\pi i \tau_x \bondop{K}} \, e^{i\Phi_y \bondop{C}},
\end{align}
where $\bondop{C}$ and $\bondop{K}$ are the conserved particle number and momentum of the Schmidt states, respectively, and $N^e$ is the total particle number. 
The first factor enforces fermion statistics of the orbitals,
	the second factor adds a $2\pi\tau_x$ twist when connecting the ends of a cylinder,
	and the final factor arises from the flux $\Phi_y$ threaded through the $y$-direction.
Via this construction, we obtain the set of MESs on a torus $\ket{\Xi_a}$ for arbitrary $\tau$ and $\Phi_{x/y}$.
By adiabatically varying these parameters, we obtain the associated Berry phases characteristic of the topological order.

The Berry phase $U_T$ is calculated as $\tau_x$ goes from $0$ to $1$ \cite{Keski-Vakkuri-1993}, corresponding to a ``$\mathcal{T}$-transformation'' of the torus as shown in Fig.~\ref{fig:torus}(a).
$U_T$ is diagonal in the MES basis, and we expect it to contain two contributions [Eq.~\eqref{eq:U_exp}].
First, when acting on $\ket{\Xi_a}$, $\mathcal{T}$ causes an anyonic flux $a$ to wind once around the $x$-cycle of the torus, generating a phase $h_a - \frac{c}{24}$, where $h_a$ is the spin of $a$  and $c$ is the chiral central charge of the edge theory.
Second, shearing the bulk introduces a phase due to the universal ``Hall viscosity'' $\HallVis$ of the fluid \cite{Avron-QHViscosity95,Read-QHViscosity09}.
$U_T$ can be calculated exactly by making use of the torus MPS [Eq.~\eqref{eq:drehen_main}], and we find that the result depends only on the infinite cylinder OES [Eq.~\eqref{eq:U_exact}].
Equating the expected and exact results, for fermions we find
\begin{subequations}\label{eq:topspin}\begin{align}
	U_{T;ab}
		&= \delta_{ab} \exp\left[ 2 \pi i \left( h_a - \frac{c}{24} - \frac{\HallVis}{2\pi\hbar} L_x^2 \right) \right]	\label{eq:U_exp}\\
		&= \delta_{ab} \, e^{ 2\pi i \left( \bondop{K}_{\bar{a}} - \dxp{\bondop{K}-\bar{n}\bondop{C}} - \nu/24 - \frac{\nu L_x^2}{16 \pi^2 \ell_B^2} \right) }. \label{eq:U_exact}
\end{align} \end{subequations}
In Fig.~\ref{fig:25topo}(b), we use the $\nu = 2/5$ OES obtained from iDMRG to extract $h$, $c$ and $\HallVis$ and find good agreement with the expected values for the Abelian hierarchy state \cite{WenZee-1992}.
For the MR phase, we obtain excellent results for the model Hamiltonian (Supplemental material \cite{Note1}), but when using only the two-body $V_m$, the measurement is highly sensitive to tunneling between the Pfaffian and anti-Pfaffian states present at the sizes studied \cite{Levin-aPf, Lee-aPf}.

\paragraph{Conclusions.}
In this paper we showed how to numerically calculate several quantities that characterize topological order starting from a microscopic FQH Hamiltonian.
The approach consists of two key steps:
	(i) We find the MPSs representation of a complete set of GSs using an iDMRG algorithm;
	(ii) we derive expressions for the QP charges, topological spins, chiral central charge, and Hall viscosity of the phase from the MPS representation.

\paragraph{Acknowledgements.}
We thank L.~Fidkowski, T.~Grover, S.~Parameswaran, A.~Turner, N.~Read, and J.~Moore for useful discussions.
We are grateful to J. B{\'a}r{\protect\fontencoding{U}\fontfamily{ipa}\selectfont\symbol{19}}arson for comments on the manuscript.
M.P.Z. acknowledges the hospitality of the guest program of MPI-PKS Dresden and support from NSF GRFP Grant No.~DGE 1106400.
R.M. is supported by NSF Grant No.~DMR-0804413 and the Sherman Fairchild Foundation.

Independently, Cincio and Vidal have developed a similar technique for using DMRG to probe quasiparticles \cite{CincioVidal}.
Also, Tu, Zhang, and Qi reported a similar method for extracting the topological spin from entanglement~\cite{TuZhangQi}.

\bibliography{qh_idmrg}
\def\qhsubputtableofcontents{\relax}
\include{qh_sup}


\end{document}

%% file: qh_sup.tex

\ifdefined\qhsubputtableofcontents \tableofcontents \else {} \fi

\begin{widetext}
\section{\texorpdfstring
	{Orbital $n$ and bond $\bar{n}$ notations}
	{Orbital n and bond bar(n) notations}}
In the subsequent analysis, sites are labeled by $n$, and bonds by $\bar{n}$.
The labels $n$ are understood to take on a numerical value indicating the location of the orbital, in units of $\frac{2 \pi \ell_B^2}{L}$.
As will be discussed in Eq.~\eqref{eq:cyl_orb}, when a flux $\Phi_x$ threads through the cylinder, $n \in \mathbb{Z} + \frac{\Phi_x}{2 \pi}$.
We let $\bar{n}$ take on a numerical value which is the average of the site locations to the bond's left and right, i.e. $\bar{n} \in \mathbb{Z} + \frac{\Phi_x + \pi}{2\pi}$.
For example, the bond between sites $n=0$ and $n=1$ is labeled by $\bar{n} = \tfrac12$.

\section{\texorpdfstring
	{Symmetry properties of Quantum Hall \lowercase{i}MPS}
	{Symmetry properties of Quantum Hall iMPS}}
\label{sec:sym}	
	As the symmetry properties of MPS are a crucial part of the details that follow, we present the symmetries of quantum Hall systems on a cylinder, a review of $\mathrm{U}(1)$ symmetry conservation for MPS, and elaborate on the $\braket{\bondop{Q}}$, $\dxp{\bondop{Q}}$ notation.

\subsection{Conserved quantities of the infinite cylinder}
	In addition to the conservation of number, it is essential to implement  conservation of momentum (also called `center of mass'), both because of the tremendous reduction in numerical effort and for the ability to label the entanglement spectrum by momentum. At filling factor $\nu$, we define the charges $(C, K)$ to be
\begin{subequations}\begin{align}
	\hat{C} &= \sum_n \hat{C}_n,
		\quad\quad \hat{C}_n = \hat{N}_n - \nu
		\quad\quad \textrm{(particle number)}	, \\
	\hat{K} &= \sum_n \hat{K}_n,
		\quad\quad \hat{K}_n = n ( \hat{N}_n - \nu )
		\quad\quad \textrm{(momentum)}	,
	\label{eq:app_charges}
\end{align}\end{subequations}
where $\hat{N}_n$ is the on-site number operator. Unlike the charge $C$, the momentum $K$ is peculiar as it behaves nontrivially under a translation by $n$ sites, $\hat{T}_n$,	
\begin{equation}
\label{TCM}
\hat{T}_{-n} \hat{K} \hat{T}_n = \hat{K} + n \hat{C}
\end{equation}
which must be accounted for when conserving the two charges in the DMRG algorithm.
If the state has a unit cell of $M$, the \emph{matrices} of the MPS are  periodic in $M$; however, due to Eq.~\eqref{TCM}, the \emph{charges} assigned to the Schmidt states are not.
Under translation, the charges of the Schmidt states on bonds $\bar{n}, \bar{n} + M$ are related by
\begin{align}
	\label{auxtcm}
	\big( \bondop{C}_{\overline{n}+M}, \bondop{K}_{\overline{n}+M} \big)
		= \big( \bondop{C}_{\bar{n}}, \bondop{K}_{\bar{n}} + M \bondop{C}_{\bar{n}} - M \dxp{\bondop{C}} \big).
\end{align}
The value $\dxp{\bondop{C}}$ is a constant of the MPS defined in the next section [Eq.~\eqref{eq:dxp}].

\subsection{\texorpdfstring
	{$\mathrm{U}(1)$ charge conservation for iMPS}
	{U(1) charge conservation for iMPS}}	
\label{sec:U1}	
Let $\hat{Q}_T = \sum_n \hat{Q}_n$ be an Abelian charge given by a sum of all  single site terms $\hat{Q}_n$.
Making a cut on bond $\bar{n}$, we can decompose $\hat{Q}_T = \sum_{n < \bar{n}} \hat{Q}_n +  \sum_{n > \bar{n}} \hat{Q}_n = \hat{Q}_L + \hat{Q}_R$.
For a \emph{finite} size state $\ket{\psi}$ with a Schmidt decomposition $\ket{\psi} = \sum_\alpha \lambda_\alpha \ket{\alpha}_L \ket{\alpha}_R$ on bond $\bar{n}$, the Schmidt states must have definite charge, $\hat{Q}_L \ket{\alpha}_L = \bondop{Q}_{\bar{n}; \alpha } \ket{\alpha}_L$.
In the MPS representation, charge conservation is expressed through the corresponding constraint \cite{Schollwock2011,SinghVidal-2011}
\begin{align}
	\left[\bondop{Q}_{\overline{r}; \beta} - \bondop{Q}_{\overline{l}; \alpha} - \hat{Q}_{n; j} \right] B^{[n] j}_{\alpha \beta} = 0
	\label{eq:Qrep}
\end{align}
where $\overline{r}/\overline{l}$ denote the bonds to the right/left of site $n$. Pictorially, exponentiating the constraint implies
\begin{align}
	\raisebox{5mm}{ \xymatrix @!0 @M=0.3mm @R=9mm @C=12mm{
		\ar[r] & {\square} \ar[r] & \\ & \ar[u]|{\lhd}_{ \displaystyle\, e^{i\theta \hat{Q}} }
	} }
	\;\;=\;\;
	\raisebox{5mm}{ \xymatrix @!0 @M=0.3mm @R=9mm @C=12mm{
		\ar[r]|{\bigtriangledown}^{ \displaystyle e^{-i\theta\bondop{Q}} }
		& {\square} \ar[r]|{\bigtriangleup}^>{ \displaystyle e^{i\theta \bondop{Q}} } & \\ & \ar[u]
	} }	.
	\label{eq:Q_pic}
\end{align}

We can define a diagonal operator $\bondop{Q}_{\bar{n}}$ acting on the auxiliary bonds of the MPS with diagonal entries $\bondop{Q}_{\bar{n}; \alpha}$.
It is convenient to define a `bond' expectation values of a $\bar{Q}_{\bar{n}}$ by
\begin{align}
\label{eq:bond_exp}
\langle \bar{Q}_{\bar{n}} \rangle \equiv \sum_{\alpha} \lambda^2_{\bar{n}; \alpha} \bondop{Q}_{\bar{n}; \alpha}
\end{align}
where $\lambda_{\bar{n}}$ are the Schmidt values on the bond $\bar{n}$. In the finite case, this gives the expected charge to the left of the cut, $\langle \bar{Q}_{\bar{n}} \rangle = \langle \hat{Q}_L \rangle$.

	In the case of an iMPS, the necessary and sufficient condition for the iMPS to have definite charge is again that there exist bond operators $\bondop{Q}_{\bar{n}}$ such that Eq.~\eqref{eq:Qrep} is satisfied.
However, Eq.~\eqref{eq:Qrep} is clearly invariant under a uniform shift of the bond charges, $\bondop{Q}_{\bar{n}} \to \bondop{Q}_{\bar{n}} + c$, so we can't obviously interpret  $\bondop{Q}_{\bar{n}}$ as the physical charges of the Schmidt states.
This ambiguity was absent in the finite case because the left-most `bond' at the boundary can canonically be assigned charge 0.
To resolve this ambiguity, we can explicitly calculate the charge of a Schmidt state using a regulator, $Q_{\alpha L} = \lim_{\epsilon \to 0} \sum_{m < \bar{n}} e^{\epsilon m} \hat{Q}_{m} \ket{\alpha}_L$.
Using Eq.~\eqref{eq:Qrep} we can rewrite this charge using that auxiliary operators $\bondop{Q}_{\bar{m}}$.
Through an abuse of notation, we write $\bondop{Q}_{\bar{m}} \ket{\alpha}_L$ as an operation on the state, by which we mean we insert $\bondop{Q}_{\bar{m}}$ into the corresponding bond of the iMPS.
We find
\begin{align}
	Q_{\alpha L} \ket{\alpha}_L	 &= \lim_{\epsilon \to 0} \sum_{m < \bar{n}} e^{\epsilon m} (\bondop{Q}_{m+1/2} - \bondop{Q}_{m-1/2}) \ket{\alpha}_L  \quad \text{[Eq.~\eqref{eq:Qrep}]}
	\notag\\
		&= \left[ \hat{Q}_{\bar{n}} - \lim_{\epsilon \to 0}  \epsilon \sum_{\bar{m} < \bar{n}} e^{ \epsilon \bar{m} }  \bondop{Q}_{\bar{m}}  \right] \ket{\alpha}_L
\end{align} 
Because $\epsilon \to 0$, the second contribution is independent of an arbitrary number of sites near the boundary $\bar{n}$. Assuming the state has finite correlation length, far from $\bar{n}$ the result becomes independent of the choice of Schmidt state $\alpha$. Making use of translation invariance (assuming a unit cell of length $M$) and the limit $\epsilon \to 0$ we find 
\begin{align}
\lim_{\epsilon \to 0}  \epsilon \sum_{\bar{m} < \bar{n}} e^{ \epsilon \bar{m} }  \bondop{Q}_{\bar{m}} \ket{\alpha}_L  = \frac{1}{M} \sum_{0 < \bar{m} < M } \langle \bondop{Q}_{\bar{m}} \rangle \ket{\alpha}_L.
\end{align}
where  $\braket{ \bondop{Q}_{\bar{m}} }$ is again the bond expectation value of the \emph{infinite} MPS.
Hence the \emph{physical} charge of the Schmidt state is
\begin{align}
	Q_{\alpha L} &= \bondop{Q}_{\bar{n}; \alpha} - \frac{1}{M} \sum_{0 < \bar{m} < M } \langle \bondop{Q}_{\bar{m}} \rangle \equiv \bondop{Q}_{\bar{n}; \alpha} - \dxp{\bondop{Q}},
\end{align}
where define the $\dxp{}$ notation:
\begin{align}
	\dxp{\bondop{Q}} \equiv& \frac{1}{M} \sum_{0 < \bar{m} < M } \braket{ \bondop{Q}_{\bar{m}} }.
	\label{eq:dxp}
\end{align}
The second term $\dxp{\bondop{Q}}$ is an average of $\langle \bondop{Q}_{\bar{m}} \rangle$ across the $M$ sites of the unit cell.
Using the freedom to shift $\bar{Q} \to \bar{Q} + c$, we can always cancel the second term $\dxp{\bondop{Q}}$ so that $\bar{Q}$ has its `naive' interpretation as the charge of the left Schmidt state.
We will call this the `canonical' choice of bond charges.
The state has fractional charges if the required constant $c$ is a fraction of the elementary charge.

\section{Numerical Methods}

	In the following we explain in detail three algorithmic issues particular to FQH DMRG on an infinite cylinder: construction of the matrix product operator (MPO) for the Hamiltonian, $\mathrm{U}(1)$ charge conservation for the momentum $K$ around the cylinder, and ergodicity issues of the DMRG update. 
The MPO formulation of iDMRG used here is explained in Refs.~\onlinecite{McCulloch-2008, Kjall-2013}.

\subsection{MPO representation of the FQH Hamiltonian}
\label{sec:numMPO}

	An MPO is a direct generalization of an MPS to the space of operators \cite{Verstraete-2004,McCulloch-2007,Kjall-2013},
\begin{equation}
	\hat{H} = \sum_{0 \leq \alpha_i < D} \cdots \otimes \hat{W}^{[1]}_{\alpha_1 \alpha_2} \otimes \hat{W}^{[2]}_{\alpha_2 \alpha_3} \otimes \cdots \, \, \, .
\end{equation}	
Each $\hat{W}^{[n]}$ is a matrix of operators acting on site $n$. The dimension $D$ of the matrices depends on the Hamiltonian, increasing as longer range components are added. For the systems studied here, $D\sim 100 \mbox{-} 300$.
An arbitrary two-body interaction in the LLL takes the form 
\begin{align}
	\hat{H} = \sum_i \sum_{0 \leq m, n} U_{mn} \psi^{\vphantom{\dag}}_{i+ 2m + n} \psi^{\dag}_{i + m + n} \psi^\dagger_{i+m}  \psi^{\vphantom{\dag}}_i + h.c.
		\owns \sum_i \sum_{0 < m, n}^\Lambda U_{m n} \psi_{i + 2m + n}^{\vphantom{\dag}} \psi^\dagger_{i + m + n} \psi^\dagger_{i + m} \psi_i^{\vphantom{\dag}}	.
	\label{happ}
\end{align}
To represent the MPO exactly in the limit of an infinite cylinder would require taking $D \to \infty$, but as the terms in the Hamiltonian of Eq.~\eqref{happ} decay like Gaussians when insertions become far apart, it is reasonable to truncate the Hamiltonian at some fixed accuracy by keeping only the largest terms. For simplicity, we will assume here a cutoff $m, n < \Lambda \sim L$.

\begin{figure}[t]
	\includegraphics[width=0.43\textwidth]{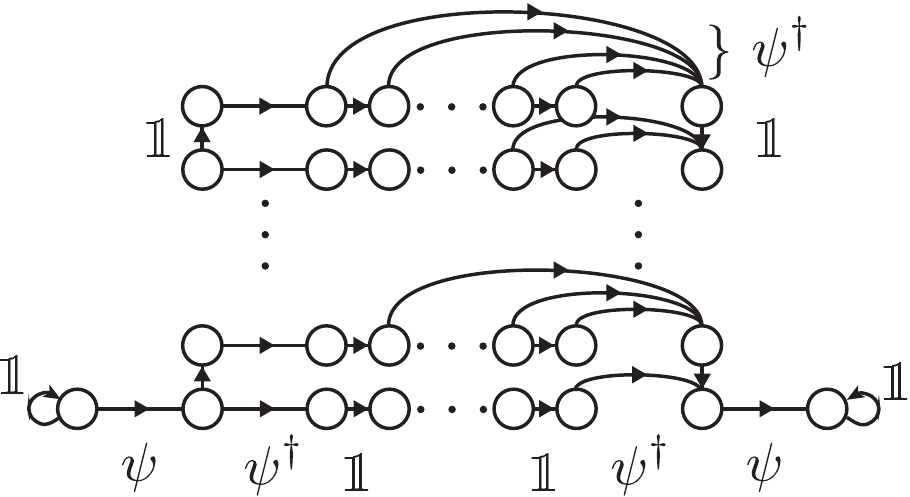}
	\caption{%
		Illustration of the finite state machine associated with the MPO representation of terms $U_{mn}$ in Eq.~\eqref{happ}.
		The machine begins in the node on the far left.
		When the first operator $\psi_i$ is placed the machine enters the square grid.
		Each row of the grid corresponds to a value of $m$; each column, a value of $n$.
		The machine proceeds vertically up the left-most column until placing $\psi^\dagger_{i + m}$, then proceeds through row $m$ until $\psi^\dagger_{i + m + n}$ is placed with weight $U_{mn}$, at which point it skips to the rightmost column.
		After descending down the rightmost column the final insertion $\psi_{i + 2m + n}$ is placed, at which point the term is complete and the machine remains in the terminating node to the far right.
		The set of all routes from the far left to far right nodes generates precisely the terms $U_{mn}$. 
		A copy, with $\psi$ and $\psi^\dag$ swapped, also exists to generate the Hermitian conjugate, as well as terms for special cases $m, n = 0$.
	}
	\label{fig:mpo} 
\end{figure}		
	To illustrate how to construct the MPO for Eq.~\eqref{happ} we view the MPO as a `finite state machine' for constructing Hamiltonians \cite{CrosswhiteBacon2008}.
Assuming translation invariance, to each index $\alpha$ of the matrix $\hat{W}_{\alpha \beta}$ we associate a state `$\alpha$' in a finite state machine, illustrated by a node in a graph.
Each non-zero entry $\hat{W}_{\alpha \beta}$ is a transition probability in the finite state machine, illustrated with an edge. 
At the $n$th step, if the machine makes the transition $\beta \to \alpha$ then the operator $\hat{W}_{\alpha \beta}$ is placed at site $n$.
The machine is non-deterministic; if there are two possible transition out of the state $\beta$, then the paths are taken in superposition, which generates the sum over all terms in the Hamiltonian.

	Assuming bosons and focusing on the particular contribution highlighted in Eq.~\eqref{happ}, at each step the MPO will place one of $\mathds{1}$, $\psi$ or $\psi^\dagger$. The MPO has a set of nodes which can be organized into a square grid, essentially in correspondence with the terms $U_{mn}$, as explained and illustrated in Fig.~\ref{fig:mpo}. The rectangular nature of the graph leads to the scaling $D \sim \Lambda^2$ of the MPO.
For fermions, the Jordan-Wigner string can be accounted for by replacing $\mathds{1}$ with the string operator $(-1)^F$ where appropriate.
As for an MPS, $\mathrm{U}(1)$ charge conservation for both number and momenta can be implemented by assigning the appropriate charges $(C, K)$ to the $D$ indices of the auxiliary bond.

\subsection{Momentum conservation on an infinite cylinder}
\label{sec:numK}

Knowing the charges transform as Eq.~\eqref{auxtcm}, the algorithm proceeds as follows.
We store the $M$ $B$-matrices of sites $n = 0, \dots, M - 1$, and keep track of the auxiliary quantum numbers on the bonds to their right,
	$\bar{n} = \frac12, \frac32, \dots, M-\frac12$.
For updates within the unit cell, charge conservation is implemented as usual.
However, for an update acting on sites $0$ and $M-1$, we must `translate' the charge data associated with site-0 to site-$M$ using Eq.~\eqref{auxtcm}.
After updating $B$, the new charge data is translated back to site 0.

\subsection{\texorpdfstring
	{Ergodicity of the \lowercase{i}DMRG algorithm}
	{Ergodicity of the iDMRG algorithm}}

	The final peculiarity of applying iDMRG to the QH effect concerns the `ergodicity' of the 2-site update.
The standard iDMRG algorithm optimizes two neighboring $B$-matrices per step in order to avoid getting stuck in local minima of the energy landscape \cite{McCulloch-2008, White-1992, Kjall-2013}.
This has the added advantage that, unlike a naive 1-site update, the bond dimension $\chi$ can grow during the simulation.
For most Hamiltonians the 2-site update is sufficient to find the optimal state, even if the initial state is taken to be a product state.
However, due to the additional constraint of momentum conservation for QH, starting from some particular state it is \emph{impossible} for the 2-site update to generate amplitude in all possible configurations.
The most naive explanation is that the smallest move available to the DMRG is a `squeeze' involving 4 sites, though this picture is not quite exact.

For the case of fermions, we have formalized and proven the following bound.
Recall that because $(C, K)$ are good quantum numbers, each Schmidt state $\alpha$ on bond $\bar{n}$ can be assigned a definite quantum number $(\bondop{C}_{\bar{n}; \alpha}, \bondop{K}_{\bar{n}; \alpha})$.
We define a combination $\bondop{P}$ of these charges by
\begin{align}
	\bondop{P}_{\bar{n}; \alpha} \equiv \bondop{K}_{\bar{n}; \alpha} - \frac{1}{2 \nu} \bondop{C}_{\bar{n}; \alpha}^2 - \bar{n}\bondop{C}_{\bar{n}; \alpha}.
	\quad\quad\quad\text{(assume $\dxp{\bondop{C}}=0$)}
	\label{eq:defP}
\end{align}	
$\bondop{P}$ has been defined so as to be invariant under translation, unlike $\bondop{K}$. For the Laughlin states, which have an entanglement spectrum in one-to-one correspondence with a chiral CFT, $\bondop{P}$ is precisely the total momentum of the CFT's oscillator modes \cite{Zaletel-2012}.
Let $\{P\}$ be the set of $\bondop{P}$'s present on all bonds in the MPS, and let $P_\text{min}, P_\text{max}$ be the minimum and maximum values they take before beginning of DMRG.
Using the standard 2-site DMRG update, $\{P\}$ always remains bounded by $P_\text{min}$ and $P_\text{max}$.
Hence the entanglement spectrum will appear to have a momentum cutoff set by the initial state, and the 2-site update will fail to find a variationally optimal state.
For example, if we use the exact Laughlin state as the seed for DMRG (which has $P_\text{min} = 0$, as the model state is purely `squeezed'), the 2-site update will not arrive at the ground state of the Coulomb Hamiltonian, which has $P_\text{min} < 0$.
		
	Though not stated in these terms, to our knowledge this ergodicity problem was previously dealt with via two methods. One approach initialized the DMRG with a large spectrum of random initial fluctuations, and ensured by hand that the  DMRG update preserves several states in each charge sector (e.g.~Ref.~\onlinecite{Zhao-2011}).
The algorithm was nevertheless observed to get stuck for several sweeps at a time, but did converge.
In this approach care is required to ensure the initial state supplies adequate fluctuations $P_\text{min}, P_\text{max}$, and there may be additional more subtle restrictions missed by this bound.
A second approach used `density matrix corrections' (e.g. Ref.~\cite{White2005, Feiguin-2008}). 

	In the current work we take a brute force (but fail safe) approach by generalizing the DMRG to a $n$-site update, optimizing $n$ sites as a time. 
In the MPS/MPO formulation of DMRG we implement a $2n$-site update by grouping $n$ sites together into a single site of dimension $d = 2^n$ and then perform the usual 2-site update on these grouped sites.
For $n=2$, for example, we simply contract 2 adjacent $B$-matrices of the MPS, taking $B^{[0]j_0} B^{[1]j_1} \to B^{[\tilde{0}] j_0 j_1}$, and likewise for the $W$-matrices of the MPO.
As the complexity of the DMRG update scales as $d^3$, this does come at a cost, though it is partially offset by the increased speed of convergence. We have not seen the algorithm get stuck while using a sufficiently expanded update.
Though we have not proved it, we believe that a $q+3$-site update will be ergodic for filling fractions $\frac{1}{q}, 1 - \frac{1}{q}$.
For more complicated fractions, such as $\frac{2}{5}$, we have checked ergodicity by trial and error; for instance, a 10-site update arrived at the same final state as a 6-site update, so the latter was used in the reported simulations.
	
	The main advantage of previous approaches, which use a two site update, is the decreased memory required, which quickly becomes a limitation for the 6-site update when $\chi \gtrsim 4000$. 
The density matrix correction approach is potentially the optimal way to proceed, once adapted to the infinite DMRG algorithm with a long range MPO, and is being developed for future work.

\subsection{Convergence with respect to truncation errors}
	The FQH iDMRG algorithm introduces two truncation errors; an error due to the finite number of Schmidt states kept, and an error due to the finite number of terms kept in the Hamiltonian (which is necessary for the MPO to have finite bond dimension). We refer to these as the MPS and MPO truncation errors respectively, which we address in turn.
	
\subsubsection{MPS truncation error}
	The MPS ansatz implies that only a finite number of states are kept in the Schmidt decomposition, which bounds the possible overlap between the MPS and the true ground state.
While the truncation relative to the exact ground state is not accessible, it is customary in iDMRG to define the `truncation error' $\epsilon_\text{MPS}$ to be the weight of the Schmidt states dropped when projecting the variationally optimal $2n$-site wave function back to the desired bond dimension.

The truncation error was kept constant while the circumference $L$ was scaled.
To simulate the system at fixed truncation error, the bond dimension grows with the circumference $L$ as \cite{Zaletel-2012}
\begin{align}
\chi \sim b e^{v c L - d_a}
\end{align}
where $c$ is the central charge of the orbital entanglement spectrum, $d_a$ is the topological entanglement entropy, and `$v$' is a non-universal number expected to vary inversely with the correlation length.
$b$ is determined by the chosen error $\epsilon$.
For the data presented in the text, $\epsilon_\text{MPS} = 10^{-5}$.
To assess the effect of finite $\epsilon_\text{MPS}$ on the extracted entanglement entropy $\gamma$, we have calculated $S(L; \epsilon_\text{MPS})$ for $\epsilon_\text{MPS} = 10^{-4}, 10^{-5}, 10^{-5.5}$.
The scaling of $S$, and the resulting estimate of $\gamma$, are reported in Fig.~\ref{fig:t_trunc}.
While $\epsilon_\text{MPS} = 10^{-4}$ introduces spurious oscillations, the relative error in $S$ between $10^{-5}$ and $10^{-5.5}$ remains about 0.2 \% (while the memory requirement is multiplied by about 1.5). 
A similar analysis must be made on a case by case basis in order to assess the tradeoff between the reduced accuracy in $S$ and the larger accessible $L$.

We note that in the most naive analysis, the truncation error  contributes an error in the topological entanglement entropy, rather than the coefficient of the area law. This may be the source of the 3\% error in out estimate of $\gamma$ for the $\nu = \tfrac{2}{5}$ state. It would be worth investigating (both for FQH an other 2D DMRG studies) whether letting $\epsilon_\text{MPS}$ scale with the circumference might remove this error.

\begin{figure}[t]
	\includegraphics[width=18cm]{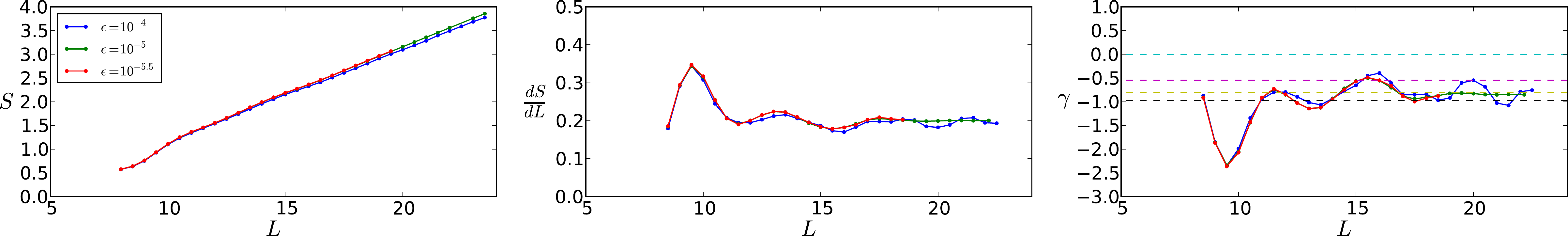}
	\caption{%
		Effect of finite $\epsilon_\text{MPS}$ on $S$ (data for $\epsilon_\text{MPS} = 10^{-5.5}$ does not extend to $L = 23.5$).
		The relative error $\Delta S/ S$ between different $\epsilon_\text{MPS}$ remains bounded.
		While large truncation error $\epsilon_\text{MPS} = 10^{-4}$ introduces spurious oscillations, $\epsilon_\text{MPS} = 10^{-5}$ and $\epsilon_\text{MPS} = 10^{-5.5}$ give the same $\gamma$ to within several percent.
	}
	\label{fig:t_trunc}
\end{figure}	

\subsubsection{MPO truncation error}
	Truncation of the Hamiltonian to a finite range smears out the interaction along the direction $y$ of the cylinder, but does preserve its locality.
To quantify the truncation error, recall that we keep only a finite number of the $V_{km}$ (say, the set $km \in A$), so we define the truncation error as $1 - \epsilon_\text{MPO} = ( \sum_{km \in A} |V_{km}| ) / ( \sum_{km} |V_{km}| )$.
We hold $\epsilon_\text{MPO} \sim 10^{-2} \mbox{-} 10^{-3}$ constant as we scale the circumference of the cylinder.
Because the spatial extent of this smearing is held constant as the circumference $L$ is increased, it is as if a fixed `cutoff' has been introduced to the Hamiltonian. When scaling $S(L) = \alpha L - \gamma$, the coefficient of the area law may modified, but the topological entanglement entropy should not be.

To illustrate the effect of the truncated Hamiltonian, we consider the model Hamiltonian for the $\nu = \frac{1}{3}$ Laughlin state.
The entanglement spectrum of the model wave function is known to have identical counting as the edge CFT; by truncating the model Hamiltonian, the entanglement spectrum is modified at large momenta.
As illustrated in Fig.~\ref{fig:mpo_trunc}, a finite `entanglement gap' is introduced with a magnitude that increases as $\epsilon_\text{MPO} \to 0$.

\begin{figure}[t]
	\includegraphics[width=17cm]{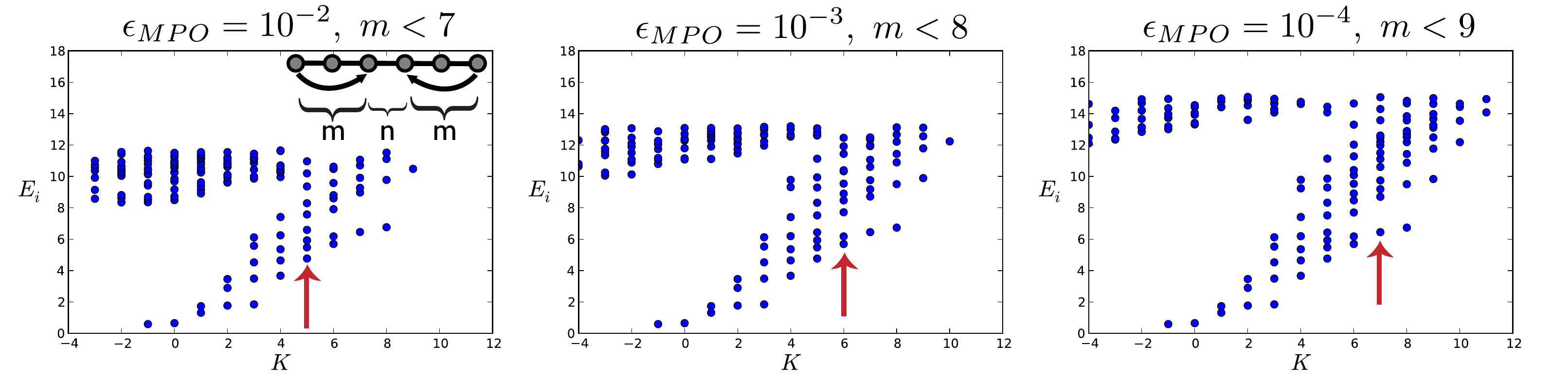}
	\caption{
		Effect of finite $\epsilon_\text{MPO}$ on the ground state entanglement spectrum of the $\nu = \frac{1}{3}$ Laughlin state at $L = 16 \ell_B$.
		The Hamiltonian approximates the `model' Hamiltonian $V_1$, but is truncated to a finite number of terms.
		As illustrated in the inset, this induces a cutoff in the squeezing distance $m \leq m_{\Lambda}$. 
		It appears that the entanglement gap intersects the spectrum at $P_{CFT} = m_{\Lambda}$, as indicated by the arrows.	
	}
	\label{fig:mpo_trunc}
\end{figure}	

\subsection{\texorpdfstring
	{Obtaining the full set of minimal entangled ground states $\{\ket{\Xi_a}\}$ from \lowercase{i}DMRG}
	{Obtaining the full set of minimal entangled ground states from iDMRG}}
	There are two issues when constructing the MES basis.
a) Does the DMRG converge to MESs, or is it favorable for it to converge to superpositions of them?
b) If it does produce MES states, how do you initialize the DMRG in order to obtain all of them? 

	As to a), it has been shown that the MES basis is in fact the eigenstate basis on an infinite cylinder, with energy densities that have an exponentially small splitting at finite circumference $L$  \cite{Cincio-2012}.
As there is no energetic reason for the DMRG to produce superpositions of the MES (and at finite $\chi$, there is in fact a finite entanglement bias to produce MES), we expect the DMRG to produce MES. 
	  
	As to b), we first consider the role of the momentum $K$ per unit cell.
The MESs are  eigenstates of the momentum $K$: if two infinite cylinder ground states with different momenta per unit cell are added in superposition, then the entanglement entropy must increase as their Schmidt states are respectively orthogonal.
The DMRG preserves $K$; so if the DMRG is initialized using a state in a momentum sector $K$ that contains an MES, then the DMRG will produce an MES; if the sector does not contain an MES, the optimized energy will observed to be higher than those of sectors that do.
The first step, then, is choose an orbital configuration $\lambda$ of the desired $K$ (such as $\lambda = 010$), initialize the iDMRG with the corresponding $\chi = 1$ MPS $\ket{\lambda}_0$ by updating the iDMRG `environments' without further optimizing the state, and then run iDMRG to obtain an optimized state and energy $E_{\lambda}$. Repeating for orbital configuration $\lambda$ of different $K$, we compare the energies $E_\lambda$ to determine which $K$ sectors contain a MES, rejecting those sectors such that $E_{\lambda}$ is not minimal within some tolerance set by the exponentially small splitting due to $L$.

For many of the expected phases (such at the Laughlin and hierarchy states), the ground-states are uniquely distinguished by $K$, so all MES will be obtained by the procedure just outlined.
For certain cases, however, several of the MES have the same $K$, for instance those corresponding to root configurations 01110 and 10101 of the $k=3$ Read-Rezayi (RR) phase at $\nu=\frac35$~\cite{ReadRezayi99}.
If we initialize the DMRG with 10101, we might worry that it will tunnel into the 01110 MES during the iDMRG, either due to the exponentially small splitting of the physical energies, or because the latter state has $d_a > 1$ and hence higher entanglement.

	We argue that the iDMRG, if initialized with an approximation of one MES $b$, will not tunnel into a different MES $a$.
Consider the following three energy scales:
(\textit{1}) the exponentially small, but physical, splitting between the ground-state energy per site due to the finite circumference, $E_{0; ab} = E_{0;a} - E_{0;b}$;
(\textit{2}) the difference in the DMRG truncation error per site, $E_{\chi; ab} = E_{\chi; a} - E_{\chi; b}$, which is inherent to the MPS representation. Each of $E_{0; a/b}$ can be made arbitrarily small for sufficient $\chi$, but, at fixed $\chi$, one of the truncation errors may be larger if $d_a \neq d_b$, as the state has higher entanglement (an effect observed for the MR state);
(\textit{3}) the gap $\Delta_{c}$ for inserting a quasiparticle of the type `$c$' that would arise at a domain wall between the two states $a, b$. 

If $E_{0; ab} + E_{\chi; ab} < 0$, heuristically the iDMRG may prefer to find the $a$ state.
In this scenario, if we initialize the iDMRG with the $b$ state, it can `tunnel' into the $a$ state by inserting a $c, \bar{c}$ pair near the sites being updated.
As the iDMRG proceeds, the state `grows' by repeatedly inserting new sites at the center of the chain.
The $c, \bar{c}$ pair then get successively pushed out to the left/right of the chain, leaving the $a$ type GS in the central region, which the state eventually converges to.
Effectively, a $c/\bar{c}$ pair has been drawn out to the edges of the `infinite' cylinder, thus tunneling between ground-states, a problem the geometry is supposed to avoid.

	However, we do not expect this to happen for energetic reasons. At a given step, the DMRG can only modify the state significantly within a correlation length $\xi$ of the bond.
Hence if a $c, \bar{c}$ pair is created, they can be drawn at most $\xi$ sites apart during the first step.
The cost to tunnel into the $a$ state during the update is	
\begin{align}
	E = \Delta_{c} - ( \xi E_{0; ab} + E_{\chi; ab})
\end{align}	
$E_{0; ab}$ is exponentially small at large $L$, and $E_{\chi}$ should be very small if sufficient $\chi$ is used, while the quasiparticle gap $\Delta_c$ remains finite.
Hence the energy of the quasiparticle provides an energetic barrier for the iDMRG to tunnel between the MESs.

If the root configurations $\lambda$ are close enough to the desired MES for the purposes of the above argument, then by initializing the DMRG with different $\lambda$ of the same $K$, the DMRG should produce the corresponding orthogonal MES.
Testing successively more complicated orbitals, we can check if we have obtained a full set by summing the quantum dimensions of the states accepted thus far \cite{Cincio-2012}. We have not verified if this proves to be the case for a non-trivial case such at the $k=3$ RR state.

	If the root configurations $\lambda$ are \emph{not} sufficiently close to the MES for the purposes of the above argument, it is also possible to run iDMRG while including a bias against the MES obtained so far in order to find the additional MES.

\section{Quasiparticle charges}
\label{sec:qp_charges}
	In the main text we claimed that the quasiparticle charges $Q_a$ are determined entirely by the entanglement spectrum of the iMPS $\ket{\Xi_a}$, which we demonstrate here in detail. As discussed, we suppose the MPS takes the form of $\ket{\Xi_{\mathds{1}}}$ for $y < 0$ and $\ket{\Xi_a}$ for $y > 0$. The most general form such a state can take is
\begin{equation}
\ket{a} = \sum_{\alpha \beta} \ket{\alpha}_L \otimes \ket{\alpha \beta}_C \ket{\beta}_R
\end{equation}
where $\ket{\alpha}_L$ are the left Schmidt states of $\ket{\Xi_{\mathds{1}}}$, $\ket{\beta}_R$ are the right Schmidt states of $\ket{\Xi_a}$, and $\ket{\alpha \beta}_C$ is an arbitrary set of states in the central `gluing' region. Without loss of generality, we suppose that the the gluing region has a length which is a multiple of $q$,  $n = \{0, 1, \cdots, q l - 1\}$ for $l \in \mathbb{Z}$. The boundary bonds $\bar{n}_L = \bar{\mathds{1}}$, $\bar{n}_R = \bar{a}$ are indexed by $\alpha/\beta$ respectively. Schematically, we should think of the low-lying Schmidt states on the left/right bond as being in the CFT sector $\mathcal{V}_{\mathds{1}}/\mathcal{V}_{a}$ respectively.

We exploit three basic facts. First, because the central region contains a multiple of $q$ sites, the charge $C$ of $\ket{\alpha \beta}_C$ must be a multiple of the electron charge (which has been chosen to be 1) for all $\alpha, \beta$. Second, the charges of the left Schmidt states are $\bondop{C}_{\bar{\mathds{1}}} - \dxp{\bondop{C}}_{\mathds{1}}$, which differ from each other only by multiples of the electron charge. Likewise, the charges of the right Schmidt states are $- (\bondop{C}_{\bar{a}} - \dxp{\bondop{C}}_{a})$, which differ from each other  only by multiples of the electron charge. Hence the charge of the quasiparticle, $Q_a$, satisfies
\begin{align}
e^{2 \pi i Q_a } = e^{2 \pi i \left[ (\bondop{C}_{\bar{\mathds{1}}} - \dxp{\bondop{C}}_{\mathds{1}}) - (\bondop{C}_{\bar{a}} - \dxp{\bondop{C}}_{a}).\right]} 
\end{align}
Again, we have taken advantage of the fact that $\bondop{C}_{\bar{n}}$ is a constant modulo 1, the charge of an electron.
Finally, because charge conjugation $\mathcal{C}$ acts as spatial inversion on the LLL orbitals, we must have $e^{2 \pi i  (\bondop{C}_{\bar{\mathds{1}}} - \dxp{\bondop{C}}_{\mathds{1}})} = 1$.
As claimed, 
\begin{align}
	e^{2 \pi i Q_a } = e^{2 \pi i \big(\dxp{\bondop{C}}_{a} - \bondop{C}_{\bar{a}}\big)} 	.
	\label{eq:Q_a}
\end{align}
When cutting a MES on any bond, $\bondop{C} \bmod 1$ is single-valued, so the equation has no ambiguity.

Note that we have implicitly assumed the sector $\mathds{1}$ appeared in the OES -- but this need not be the case, as for instance in the $\nu = \frac{1}{2}$ bosonic Laughlin state.
If we naively apply the above formula to $\nu = \frac{1}{2}$  Laughlin, we find particles of charge $\pm e/4$, rather than $0, e/2$. 
This issue is clarified using an alternate derivation of the charge via \hyperref[sec:flux_mat]{flux matrices}.

\section{Matrix product states on a torus}
\newcommand{\phicyl}{\varphi^\text{cyl}}	

Here we explain in detail the procedure to take an infinite cylinder iMPS to a torus MPS.
Our approach is closely related to that of Ref.~\onlinecite{Cincio-2012}, but with the additional complication of having twisted boundary conditions.

\subsection{From an infinite chain to a periodic one}
	We first step back and show how to convert the MPS of any gapped, infinite chain to a periodic chain, both for the trivial case when there is no `twist,' and then in the presence of a twist generated by a symmetry.
The construction in the first case, which we explain for completeness, is obvious: we cut out a segment of the iMPS and reconnect the two dangling bonds to form a ring.
For simplicity in what follows we will assume bosonic chains, and introduce the correction for Fermionic chains due to the Jordan-Wigner string at a later point.	
	
Consider two systems, an infinite chain with a unit cell of $N$, and a periodic chain of length $N$.
The sites of the chains are labeled by `$n$' (with $n \sim n + N$ in the periodic case).
Restricting to local Hamiltonians, we can decompose  $\hat{H} = \sum_n \hat{H}^{[n]}$, where each $\hat{H}^{[n]}$ is localized around site $n$ over a length $\xi_H \ll N$ (for bosons, for example, these are the terms $\hat{H}^{[n]} = \hat{b}^\dagger_{n+1} \hat{b}_n + h.c.$, though could extend over many sites).
In the infinite case, translation symmetry implies $\hat{T}^{N} \hat{H}^{[n]} \hat{T}^{-N} = \hat{H}^{[n+N]}$, $\hat{T}$ being the translation operator.
	
	The energetics of the state are determined by the reduced density matrices of the system, $E = \sum_n \mbox{Tr}( \hat{\rho}^{[n]} \hat{H}^{[n]})$.
The $\hat{\rho}^{[n]}$ are reduced density matrices in a region around $n$ large enough to include all the sites affected by $\hat{H}^{[n]}$.
If the $\hat{H}^{[n]}$ of the finite and infinite chains are identical, then to find the ground state of the periodic system it will be sufficient to reproduce the local density matrices $\hat{\rho}^{[n]}$ of the infinite system. 
To do so, we first cut out a segment of the iMPS with two dangling bonds, $\Psi_{\alpha\beta} = B^{[0]} B^{[1]} \cdots B^{[N-1]}$, or in the pictorial representation of MPS:
\begin{align}
	\Psi_{\alpha\beta} \;=\;
	\raisebox{5mm}{\xymatrix @M=0.3mm @R=8mm @C=6mm @!C{
		{\alpha}	\ar@{-}[r]
		&	{\square} \ar@{-}[r] \ar@{-}[d]
		&	{\square} \ar@{--}[rr] \ar@{-}[d]
		&
		&	{\square} \ar@{-}[r] \ar@{-}[d]
		&	{\square} \ar@{-}[r] \ar@{-}[d]
		&	{\beta}
	\\	& 0 & 1 & & N-2 & N-1 &
	}}.
\end{align}
We then connect (trace over) the dangling bonds to form a ring MPS:
\begin{align}
	\label{eq:ringtrick}
	\operatorname{Tr}[\Psi]	&\;=
	\raisebox{10mm}{\xymatrix @M=0.3mm @R=7mm @C=6mm @!C{
		&& \ar@{--}[r] &&&
	\\	&	{\square} \ar@{-}[r] \ar@{-}[d]
			\ar@{-} `l[u] `u[u] [ur]
		&	{\square} \ar@{-}[r] \ar@{-}[d]
		&	{\square} \ar@{-}[r] \ar@{-}[d]
		&	{\square} \ar@{-}[d]
			\ar@{-} `r[u] `u[u] [ul] &
	\\	&	N-2 & N-1 & 0 & 1 &
	}},
\end{align}
There is no need to further optimize the MPS at the `seam,' because when the complement to the region of $\hat{\rho}^{[n]}$ is large compared to the correlation length, $N - \xi_H \gg \xi$, then up to corrections of order $e^{-N/\xi}$ the $\hat{\rho}^{[n]}$ of the periodic MPS are identical to those of the iMPS.

\subsubsection{Twists} 
We now introduce a twist generated by a local symmetry, $\hat{Q} = \sum_n \hat{Q}^{[n]}$, assuming each $\hat{H}^{[n]}$ is individually symmetric under $\hat{Q}$ and that $\hat{T}^{N} \hat{Q}^{[n]}\hat{T}^{-N} = \hat{Q}^{[n+N]}$.
We use a unitary `twist' operator $\hat{G}$ to introduce a twist every $N$ sites of infinite chain, which generates a twisted Hamiltonian:
\begin{align}
\hat{G}(\theta) =  \prod_b e^{i b  \theta  \sum_{n = 0}^{N-1} \hat{Q}^{[n+b N]} }, \quad \quad \hat{H}_{\infty}(\theta) \equiv \hat{G}(\theta) \hat{H}_\infty \hat{G}(-\theta) =  \sum_n \hat{H}^{[n]}(\theta).
\end{align}
The chain remains translation invariant, $\hat{T}^{N} \hat{H}^{[n]}(\theta) \hat{T}^{-N} = \hat{H}^{[n+N]}(\theta)$, because $\hat{Q}$ is a symmetry.

We define the twisted Hamiltonian of the periodic chain to be $\hat{H}_{\circ}(\theta) = \sum_{n=0}^{N-1} \hat{H}^{[n]}(\theta)$ with $\hat{H}^{[n]}(\theta)$ taken from the infinite chain.
In the bosonic case, for example, there is a single link with a twist, $b_{0}^\dag b_{-1} e^{i\theta} + h.c.$.
Note that $\hat{H}_{\circ}(\theta)$ is not unitarily related to  $\hat{H}_{\circ}(0)$ except when $\theta$ is a multiple of a `flux quantum' $\Phi$, which  we can assume is $\Phi = 2 \pi$ .
What is the ground state $\ket{\theta}_\circ$ of $\hat{H}_{\circ}(\theta)$? 

	By construction, the Hamiltonians $H_{\circ}, H_{\infty}$ remain locally identical, so given the iMPS for $\ket{0}_\infty$, we can utilize the gluing trick of Fig.~\eqref{eq:ringtrick} already justified.
As $\ket{\theta}_\infty = \hat{G}(\theta) \ket{0}_\infty$, with $\ket{0}_\infty$ the untwisted ground state, the desired iMPS is	
\begin{align}
	\ket{\theta}_\infty &=
	\raisebox{5mm}{\xymatrix @M=0.3mm @R=8mm @C=6mm @!C{
		{\cdots\;\;}	\ar@{-}[r]
		&	{\square} \ar@{-}[r] \ar@{-}[d]
		&	{\square} \ar@{-}[r] \ar@{-}[d]
		&	{\square} \ar@{-}[r] \ar@{-}[d]|{\lhd}^{\displaystyle\;e^{i\theta\hat{Q}}}
		&	{\square} \ar@{-}[r] \ar@{-}[d]|{\lhd}^{\displaystyle\;e^{i\theta\hat{Q}}}
		&	{\square} \ar@{-}[r] \ar@{-}[d]|{\lhd}^{\displaystyle\;e^{i\theta\hat{Q}}}
		&	{\;\;\cdots}
	\\	&	-2 & -1 & 0 & 1 & 2 &
	}}
\end{align}
and so on throughout the chain. Using the conservation rule of \eqref{eq:Q_pic}, we can rewrite the above as
\begin{align}
	\label{eq:sitetobond}
	\ket{\theta}_\infty &=
	\raisebox{5mm}{\xymatrix @M=0.3mm @R=8mm @C=6mm @!C{
		{\cdots\;\;}	\ar@{-}[r]
		&	{\square} \ar@{-}[r] \ar@{-}[d]
		&	{\square} \ar@{-}[r]|{\bigtriangledown}^{ \displaystyle \; e^{-i\theta\bondop{Q}} } \ar@{-}[d]
		&	{\square} \ar@{-}[r] \ar@{-}[d]
		&	{\square} \ar@{-}[r] \ar@{-}[d]
		&	{\square} \ar@{-}[r] \ar@{-}[d]
		&	{\;\;\cdots}
	\\	&	-2 & -1 & 0 & 1 & 2 &
	}}
	.
\end{align}
To obtain $\ket{\theta}_\circ$, we again cut out a segment and glue,
\begin{align}
\label{eq:torus_mps_def}
	\ket{\theta}_\circ	\;=
	\raisebox{10mm}{\xymatrix @M=0.3mm @R=7mm @C=6mm @!C{
		&& \ar@{--}[r] &&&
	\\	&	{\square} \ar@{-}[r] \ar@{-}[d]
			\ar@{-} `l[u] `u[u] [ur]
		&	{\square} \ar@{-}[r]|{\displaystyle\bullet}^*+++{\displaystyle\drehen} \ar@{-}[d]
		&	{\square} \ar@{-}[r] \ar@{-}[d]
		&	{\square} \ar@{-}[d]
			\ar@{-} `r[u] `u[u] [ul] &
	\\	&	N-2 & N-1 & 0 & 1 &
	}}	.
\end{align}
where $\bondop{G} = e^{-i\theta\bondop{Q}}$. 

\subsubsection{Fermions}
One additional modification must be made for fermions, due to the Jordan Wigner string.
MPS for fermionic chains are always expressed through the occupation of bosonic operators $\sigma^+_n = (-1)^{\sum_{j < n} \hat{N}_j }\psi^\dagger_n$, where $\hat{N}_i$ is the occupation at site $i$.
In other words, the $B$ matrix at site $n$ generates the state according to $B^{j_n} (\sigma^+_n)^{j_n} \ket{0}$.
On a periodic chain, the fermionic operators $\psi_n \sim \psi_{n + N}$ are periodic, but because of the string the  $\sigma^+_n$ are not. If $\bondop{N}$ is the bond operator corresponding to number, we find the string can be accounted for via
\begin{align}
	\bondop{G} = \eta^{(N^F - 1) \bondop{C}} e^{-i\theta\bondop{Q}}, \quad \eta = \pm\textrm{ for Boson/Fermion}
	\label{eq:fermionsign}
\end{align}
where $N^F$ is the total fermion number and $\eta = \pm1 $ for bosons and fermions respectively.

In the following sections, we present the details for this construction pertaining to quantum Hall systems.
First we describe the single-particle basis on an infinite \hyperref[sec:orb_cyl]{cylinder} by solving the Hamiltonian $H_0 = \frac12 (\mathbf{p} + \mathbf{A})^2$, then adapt the basis to a \hyperref[sec:orb_torus]{torus}.
We then show that in the orbital basis, the cylinder and torus Hamiltonians $H_{\infty}, H_\circ$, satisfy the criteria just discussed, so we can obtain the \hyperref[sec:mps_torus]{torus MPS} from the cylinder MPS.

\subsection{The cylinder geometry}

\label{sec:orb_cyl}
Consider a cylinder finite in the $x$-direction with circumference $L_x$, infinite in the $y$-direction, with the following boundary condition and Hamiltonian:
\begin{align}
	\psi(x, y) &= \psi(x - L_x, y) e^{i\Phi_x}	,
	\label{eq:torus_bc_x}
	\\
	H_0 &= \frac12 (\mathbf{p} + \mathbf{A})^2	,
\end{align}
where $\Phi_x$ is the flux threading the $x$-cycle and $\mathbf{p} = -i\nabla$.
In the Landau gauge $\mathbf{A} = \ell_B^{-2}(-y, 0)$, the $x$-momentum $k$ is a conserved quantity.
In that eigenbasis, the wave functions $\phicyl_n(x,y) = \braket{x,y | \phicyl_n}$ in the lowest Landau levels (LLL) are
\begin{align}
\label{eq:cyl_orb}
	\phicyl_n(x,y) &= \frac{1}{\sqrt{L_x\ell_B{\sqrt{\pi}}}} e^{ik_nx} e^{-\frac{(y-y_n)^2}{2\ell_B^2}}	,
	&&\textrm{with}\;
	k_n = n \frac{2\pi}{L_x} ,\;\;
	y_n = k_n \ell_B^2 ,\;\;
	n \in \mathbb{Z} + \frac{\Phi_x}{2\pi} .
\end{align}
We see that the centers of the wave functions $y_n$ depends on $\Phi_x$; pumping a flux through $x$-loop will shift all the orbitals by unit distance $\Delta y = 2\pi\ell_B^2/L_x$.
Although the phases of these orbitals are arbitrary, it is convenient to resolve the ambiguity via the translation operator:
\begin{align}
	\hat{T}_y = \exp\left[ -i\tfrac{2\pi}{L_x} \big({-i} \ell_B^2 \partial_y - x\big) \right]	,
\end{align}
which shifts the wavefunctions by $\Delta y$.
We have chosen the phases of the orbitals such that $\ket{\phicyl_{n+1}} = \hat{T}_y \ket{\phicyl_n}$.

\textit{Note.}
We emphasize that the flux $\Phi_x$ is accounted for by labeling sites as $n \in \mathbb{Z} + \frac{\Phi_x}{2\pi}$.
This carries over to the definition of the momentum: $\hat{K}_n = n \hat{C}_n$ includes the \emph{fractional} part $\frac{\Phi_x}{2\pi}$.
This definition of $\hat{K}_n$ is an important detail for the formulas that follow.

\subsection{The torus geometry}
\label{sec:orb_torus}

To go from a cylinder to a torus, we identify points $(x,y)$ with $(x-\tau_xL_x, y-L_y)$ as follows
\begin{align}
	\psi(x, y) = \psi(x-\tau_xL_x, y-L_y) e^{i\ell_B^{-2}L_yx + i\Phi_y}	.
	\label{eq:torus_bc_y}
\end{align}
or equivalently $\psi(x, y) = \psi(x+\tau_xL_x, y+L_y) e^{-i\ell_B^{-2}L_y(x+\tau_xL_x) - i\Phi_y}$.
The phase $i\ell_B^{-2}L_yx$ is necessary since the gauge potential $\mathbf{A}$ is not periodic when taking $y \rightarrow y+L_y$,
	the difference being $\mathbf{A}(x-\tau_xL_x,y-L_y) - \mathbf{A}(x,y) = \nabla(\ell_B^{-2}L_yx)$.
Note that for the above to be well-defined, we need an integer number of fluxes in the torus:
\begin{align}
\frac{L_x L_y}{2\pi\ell_B^2} = N_\Phi \in \mathbb{Z}	.
\end{align}
The fluxes $(\Phi_x,\Phi_y)$ parameterizes a set of Hamiltonians and corresponding Hilbert space of ground states.

To figure out the eigenstates for the torus, we sum over combinations of $\phicyl_n$ on the cylinder such that they respect the boundary condition above.
The single-particle wave functions $\varphi_n(x,y) = \braket{x,y | \varphi_n}$ for the torus are
\begin{align}
\label{eq:phitor}
	\varphi_n(x,y) &= \sum_b \phicyl_{n+bN_\Phi}(x,y) \,
		e^{-2\pi iN_\Phi \frac{b(b-1)}{2} \tau_x - 2\pi ibn\tau_x + ib\Phi_y} .
\end{align}
We will use as our basis the orbitals $ 0 \leq n < N_\Phi$.


\subsection{MPS for twisted tori}
\label{sec:mps_torus}

	Since the flux $\Phi_y$ corresponds to a twist generated by the charge $\hat{C}$, while the modular parameter $\tau_x$ corresponds to a twist generated by the momentum operator $\hat{K}$, it is natural to suppose the orbital MPS can be obtained using a twist generated by $2 \pi \tau_x \hat{K} - \Phi_y \hat{C}$.
We prove this intuition is correct, but the surprise is that the ground state remains exact (up to corrections $\mathcal{O}^{-L_y/\xi}$) even though the twist is performed in \emph{orbital} space, rather than real space.
	
	We first determine the form of the orbital Hamiltonian of the torus, $\hat{H}_\circ(\theta)$, in the basis of Eq.~\eqref{eq:phitor}.
We assume the Hamiltonian arises from products of the electron density operators $\rho(\vec{x})$, is translation invariant in both $x$ and $y$, and that any interactions, such as $\rho(\vec{x})V(\vec{x} - \vec{y})\rho(\vec{x})$, are short ranged in comparison to the length $L_y$ of the torus.
As we will eventually take $L_y \to \infty$, this is not really a restriction.

	We make use of the interaction overlap integrals for the \emph{cylinder} orbitals $\varphi^{cyl}_{n_i}$, which we denote $V_{ \{n_i\}}$.
The two body term, for instance, is encoded in a term $V_{n_1 n_2 n_3 n_4}$.
$V_{\{n_i\}}$ is non-zero only if $\sum_i n_i = 0$, due to momentum conservation around the cylinder, and $V_{\{n_i + 1\}} = V_{\{n_i\}}$ due magnetic translation invariance along the cylinder.
Furthermore, $V_{ \{n_i\} }$ decays  when the separation between  indices are such that $L_x/\ell_B  \ll  |n_i - n_j|$.
Since $L_x/\ell_B \ll N_\Phi$, to exponentially good accuracy we can assume $V_{ \{n_i\} } = 0 $ if any two $n_i$ differ by $N_\Phi$ sites.

For notational simplicity we will illustrate the calculation for the 1-body term, which generalizes in an obvious fashion. 
The 1-body Hamiltonian is
\begin{align}
\hat{H} &= \sum_{ 0 \leq n_i < N_\Phi } \sum_{0 \leq b_i \leq 1} V_{n_1 + b_1 N_\Phi, n_2 + b_2 N_\Phi} e^{-f(n_1, b_1) + f(n_2, b_2)} c^\dagger_{n_1} c_{n_2} \\
f(n, b) &= -2\pi iN_\Phi \frac{b(b-1)}{2} \tau_x - 2\pi ibn\tau_x + ib\Phi_y
\end{align}
The summand is invariant under taking all $b_i \to b_i + 1$, so we can safely restrict to terms such that $b_i \in \{0, 1\}$ with at least \emph{one} of the $b_i = 0$ ( $V$ vanishes if two indices are $N_\Phi$ apart).
This leads to two types of terms: the `bulk' terms, in which all $b_i = 0$, and the `seam' terms, in which some $b_i = 0$ and some $b_i = 1$.
The seam terms arise when site at the beginning and end of the unit cell interact.

Since $f(n, 0) = 0$, in the bulk the local Hamiltonians $H^{[n]}(\theta) = H^{[n]}(0)$ are identical to the infinite cylinder Hamiltonians.
Along the seam the $H^{[n]}(\theta)$ acquire phases $f(n_i, 1) = -2 \pi i \tau_x n_i + i \Phi_y$ for each orbital with $b_i = 1$.
For example, if $N_\Phi = 10$, in the interaction $V_{9, 11} c^\dagger_9 c_1 \to V_{9, 11} c^\dagger_9 c_1 e^{f(1, 1)}$.

Consequently the torus $\hat{\rho}^{[n]}$ should be identical to the infinite cylinder $\hat{\rho}^{[n]}$ in the bulk, but differ by phases $f(n_i, 1)$ near the seam.
To account for the phase $f(n_i, 1) = -2 \pi i \tau_x n_i + i \Phi_y$, we use the same conserved quantity manipulations as we did in Eq.~\eqref{eq:sitetobond}, and find the correct twist operator $\bondop{G}$ is indeed
\begin{align}
	\drehen &= \eta^{(N^e - 1) \bondop{C}}
		\exp\left[ -2\pi i\tau_x \bondop{K} + i\Phi_y\bondop{C} \right] .
	\label{eq:drehen}
\end{align}
$\eta = \pm1$ accounts for bosons and fermions respectively, as discussed for Eq.~\eqref{eq:fermionsign}.
There is one subtlety we must emphasize: recall that $\bondop{K}$ is not periodic, due to Eq.~\eqref{TCM}, so the quantum numbers $\bondop{K}_\alpha$ according to the right bond of the last site and the left bond of the first site differ by $N_\Phi \bondop{C}$.
In the above form we assume $\bondop{K}$ uses the quantum numbers according to the bond before the first site, at $\bar{n} = \frac{\Phi_x}{2 \pi}-\frac12$, \emph{not} after the last site.
The first factor $\eta^{(N^e-1)\bondop{C}}$ reflects the fermion statistics among the orbitals.

\section{Computation of flux and modular matrices via the entanglement spectrum}
\label{sec:berry_matrix_comp}

The flux matrices $\mathcal{F}_{x,y}$ gives the action of threading a $2\pi$ flux in the $x,y$-loop, while the modular $\mathcal{T}$-matrix gives the action of a Dehn twist.
To derive their expression, we compute the (non-Abelian) Berry phase from the adiabatic changes
	$\Phi_{x,y} \rightarrow \Phi_{x,y} + 2\pi$ and $\tau_x \rightarrow \tau_x + 1$, respectively.
In each case a parameter $\kappa$ is varied and the Berry phase $U$ is a $\gsd \times \gsd$ matrix which is a product of two pieces:
\begin{align}
	U &= W \left(\mathcal{P} e^{ i \int_{\kappa_i}^{\kappa_f}\!\!d\kappa \, A }\right)	,
\end{align}
with $A$ and $W$ being matrices defined as
\begin{align}
	A_{ab}(\kappa) &= \Braket{ \Xi_b(\kappa) | -i\tfrac{\partial}{\partial\kappa} | \Xi_a(\kappa) }	,
&	W_{ab} &= \Braket{ \Xi_b(\kappa_f) | \Xi_a(\kappa_i) }	.
\end{align}
As $\kappa$ is varied, $\ket{\Xi_a(\kappa)}$ denotes the set of ground states for the boundary conditions specified by $\kappa$;
$A$ is the Berry connection and $W$ is the overlap between the initial and final set of MES.
While the result is independent of the choice of basis, in the MPS construction $ \ket{\Xi_a(\kappa_i) }$ remains the MES basis.

The action of $\mathcal{F}_{x/y}, \mathcal{T}$ are characteristic of the topological phase and robust to perturbations.
The results	presented are summarized in Tab.~\ref{tab:BerryPhases}.
Although the flux matrices and topological spins are formalized for a state in the torus geometry, we will show that these quantities only depend on the OES of an infinite cylinder.
In particular, knowing the entanglement spectrum of each MES $\ket{\Xi_a}$ allows us to determine the charge $Q_a$ and  topological spin $h_a$ of the quasiparticle $a$.

\textit{Note.}
There is a subtle point in the interpretation of the resulting equations.
The charge $Q_a$ and the spin $h_a$ we will subsequently derive are those of the sector $\mathcal{H}_a$ that would arise in the real space entanglement spectra \emph{if we were to cut the system at} $y=0$.
The entanglement spectra appearing at $y=0$ depend on $\Phi_x$.
It is most natural to choose $\Phi_x$ such that $a = \mathds{1}$ can appear on the cut at $y=0$: the resulting set of `$a$' are then those that would appear on the plane.
We find that for fermions, this occurs for $\Phi_x = \pi$, while for bosons, this occurs for $\Phi_x = 0$.
Hence it is most natural to set $\Phi_x = 0$ for bosons, and $\Phi_x = \pi$ for fermions in the following equations.

\begin{table}[h]
	\begin{tabular}{cccc}
	\hline\hline
		$\kappa$ parameter	&	Berry phase matrix	&	Physical observable	&	Formula	\\
	\hline
		$\Phi_x$	&	$\mathcal{F}_x$	&	translation structure	&	Eq.~\eqref{eq:Fx}	\\
		$\Phi_y$	&	$\mathcal{F}_y$	&	charge	&	Eq.~\eqref{eq:Fy} \\
		$\tau_x$	&	$\mathcal{T}$	&	topological spin, Hall viscosity, central charge	&	Eq.~\eqref{eq:T}	\\
	\hline\hline
	\end{tabular}
	\caption{%
		Summary of the non-Abelian phases from pumping $\Phi_x, \Phi_y$ and $\tau$.
	}
	\label{tab:BerryPhases}
\end{table}

\subsection{Computing the Berry phase}
	We have already determined the many body state $\ket{\Xi_a(\kappa)} = \Psi^a_{j_0 j_1 \cdots}(\kappa)\ket{j_0, j_1 \cdots; \kappa}$, so it is in principle a mechanical matter to calculate the Berry connection.
	The orbital wave function $\Psi^a$ is given by the periodic MPS with a twist [Eq.~\eqref{eq:torus_mps_def}], and $j_n$ is the occupation number of orbital $\varphi_n$ defined in Eq.~\eqref{eq:phitor}. 
Our approach is as follows: we first derive the general structure of the Berry phase, by decomposing the result into the `wavefunction' and `orbital' parts.
We then report the results for the constituent terms for $\mathcal{F}_{x/y}, \mathcal{T}$. 	
	
\begin{itemize}
\item{$A$. }
The connection $A_{ab}(\kappa)$ can be separated into two components, one from the changing twist in the MPS $\Psi^a$, and one from the changing orbitals:
\begin{align}
\label{eq:A_masterEq}
A_{ab}(\kappa) &= \sum_{ \{j\} }\bar{\Psi}^a_{ \{j\} } (-i \partial_\kappa)  \Psi^b_{\{j\}} +    \sum_{ \{j\}, \{k\} }\bar{\Psi}^a_{\{j\}} \Psi^b_{\{k\}} \, \bra{\{j\}; \kappa} (-i \partial_\kappa) \ket{\{k\}; \kappa} \equiv A_{ab}^{(\drehen)}(\kappa) + A_{ab}^{(\varphi)}(\kappa) 
\end{align}
The first term of Eq.~\eqref{eq:A_masterEq} involves the Berry connection for $\Psi^a$, which depends on $\kappa$ through the twist operator $\drehen$.
In the limit of $L_y\gg \xi$, the required overlap reduces to a `bond expectation value' as defined in Eq.~\eqref{eq:bond_exp},
\begin{align}
	A_{ab}^{(\drehen)} = -i \delta_{ab}\Braket{\drehen^\ast \frac{\partial\drehen}{\partial \kappa}} .
\end{align}
The expectation value is taken on the bond where $\drehen$ sits, to the left site  $n = \frac{\Phi_x}{2\pi}$.
The second term of Eq.~\eqref{eq:A_masterEq} involves the Berry connection between orbitals:
\begin{align}
	A_{ab}^{(\varphi)}(\kappa) = -i \delta_{ab} \sum_n \braket{ \hat{N}_n } \braket{\varphi_n | \partial_\kappa | \varphi_n},
	\quad   \int_{\kappa_i}^{\kappa_f}\!\! A_{ab}^{(\varphi)}(\kappa) \, d\kappa \equiv \sum_n \theta^{(\varphi)}_n \braket{ \hat{N}_n }
\end{align}
where $\braket{ \hat{N}_n }$ is the average occupation at site $n$.

\item{$W$. }
Likewise, the overlap $W_{ab}$ also contains a twist and orbital component,
\begin{align}
W_{ab}&= \sum_{ \{j\}, \{k\} }\bar{\Psi}^a_{\{j\}}(\kappa_f) \braket{\{j\}; \kappa_f | \{k\}; \kappa_i}  \Psi^b_{\{k\}}(\kappa_i).
\end{align}
We distinguish between two cases: for $\mathcal{F}_y$ and $\mathcal{T}$, the orbitals return to themselves (), giving
\begin{align}
W_{ab}= \sum_{ \{j\}}\bar{\Psi}^a_{\{j\}}(\kappa_f) \Psi^b_{\{j\}}(\kappa_i) = \delta_{ab} \frac{\drehen|_{\kappa=\kappa_i}}{\drehen|_{\kappa=\kappa_f}} \quad ( \mbox{for }\,  \mathcal{F}_y,\, \mathcal{T} ) . 
\end{align}
For $\mathcal{F}_x$, the orbitals are translated by 1 site, and we find
\begin{align}
W_{ab}&=\braket{\Xi_b | \hat{T}_y^{-1} | \Xi_a} \frac{\drehen|_{\kappa=\kappa_i}}{\drehen|_{\kappa=\kappa_f}} \quad ( \mbox{for }\, \mathcal{F}_x).
\end{align}
\end{itemize}

Because of this decomposition, the total Berry phase takes the form
\begin{align}
	U_{ab} &= W_{ab} \, \exp\left[ i\int\! A^{(\drehen)} \right]
		\exp \Braket{ i \sum_\textrm{sites $n$}\! \theta^{(\varphi)}_n \hat{N}_n }
	,  \label{eq:U_masterEq}
\end{align}
Below we summarize the results for each case, giving an explicit formula for $\mathcal{F}_x, \mathcal{F}_y$ and $\mathcal{T}$.

\subsubsection{Orbital Berry Phase} 

A technical difficulty arises when we try to compute the orbital Berry phase $\theta^{(\varphi)}_n$, due to the changing boundary conditions.
To remove all ambiguity, we follow the approach of Ref.~\cite{Keski-Vakkuri-1993} by working in a coordinate system $X, Y$ with \emph{fixed} boundary conditions (which determines the Hilbert space) and vary the metric and gauge potential, which determines the Hamiltonian. In this approach, the boundary conditions, metric, and vector potential read
\begin{subequations}\begin{align}
	\tilde\psi(X,Y) &= \tilde\psi(X+L_x, Y)	,
\\	\tilde\psi(X,Y) &= \tilde\psi(X, Y+L_y) \, e^{-i\ell_B^{-2}L_yX}	,
\\	\tilde{g}_{\mu\nu} &= \begin{pmatrix}
				1 & \dfrac{\tau_x}{\tau_y} \\ \dfrac{\tau_x}{\tau_y} & 1+\tau_x^2/\tau_y^2
			\end{pmatrix}	,
		\quad\textrm{with } \tau_y = L_y/L_x	,
\\ \tilde{A}_\mu &= \begin{pmatrix} \frac{\Phi_x}{L_x} - \ell_B^{-2}Y, & \frac{\Phi_y}{L_y} + \frac{\pi\tau_xN_\Phi}{L_y} \end{pmatrix}	.
\end{align}\end{subequations}
The odd looking term $\frac{\pi\tau_xN_\Phi}{L_y} \in \tilde{A}_Y$ exists to counteract the $N_\Phi/2$ flux quanta inserted as $\tau_x$ increases by 1.

	The coordinate system $(x,y)$ and boundary conditions of Eq.~\eqref{eq:torus_bc_y} are unitarily related to $(X, Y)$ through a change of coordinates and gauge transformation. While we omit the details of the computation, all the orbital overlaps and Berry phases can be unambiguously defined by transforming the orbitals to the $(X, Y)$ coordinates.

	We also note that under $\mathcal{T}$, examining the single particle orbitals discussed above shows that the fluxes transform as $(\Phi_x, \Phi_y) \to (\Phi_x, \Phi_y + \Phi_x)$, so $U_T$ does not take the system back to itself unless $\Phi_x = 0$. We return to this issue when discussing modular transformations.

\subsection{\texorpdfstring
	{Flux matrices $\mathcal{F}_{x/y}$: quasiparticle charge and Hall conductance}
	{Flux matrices: quasiparticle charge and Hall conductance}}
\label{sec:flux_mat}
\begin{itemize}
\item{$\mathcal{F}_y$. }
The flux matrix $\mathcal{F}_y$ describe how the MES $\ket{\Xi_a}$ transform as a flux quanta is threaded through the $y$-loop.
Letting $\kappa = \Phi_y$, the pieces in Eq.~\eqref{eq:U_masterEq} are as follows,
\begin{align}
	\theta^{(\varphi)}_n &= -2\pi\frac{n}{N_\Phi}	,
&	A^{(\drehen)} &= \braket{\bondop{C}}	,
&	W_{ab} &= \delta_{ab} e^{-2\pi i\bondop{C}_{\bar{a}}}	.
\end{align}
The number operator $\hat{N}$ may be rewritten in terms of bond operators,
	\textit{e.g.} $\hat{N}_n = \hat{C}_n + \nu = \bondop{C}_{\overline{n+1/2}} - \bondop{C}_{\overline{n-1/2}} + \nu$.
With a bit of algebraic manipulation, the Berry phase can be written as
\begin{align}
	\mathcal{F}_y &= \exp\left[ 2\pi i (\dxp{\bondop{C}}-\bondop{C} ) + i\nu(\Phi_x-\pi) \right]
	\label{eq:Fy}
\end{align}

We see that our formula for $\mathcal{F}_y$ is similar to that for the charge $e^{2\pi i Q_a}$ in Eq.~\eqref{eq:Q_a}.
Note that $\mathcal{F}_y$ is diagonal in the basis of the MES $\ket{\Xi_a}$.
Physically, this is because each quasiparticle type has a well-defined charge, modulo the electron charge.
In Eq.~\eqref{eq:Fy} we explicitly include the $\Phi_x$ dependence, whereas in Eq.~\eqref{eq:Q_a} we implicitly assumed $\Phi_x = \pi$, which is most natural for fermions.
Since $\Phi_x$ translates the state (see below), the term $\nu \Phi_x$ encodes the Su-Schrieffer counting.

\item{$\mathcal{F}_x$. }
Inserting a flux quanta through the $x$-loop will in effect translate the MPS by one unit cell.
We can see that the orbitals momenta, in units of $\frac{2\pi}{L_x}$, are quantized and of the form $n = \mathbb{Z} + \frac{\Phi_x}{2\pi}$,
	and hence as $\Phi_x$ adiabatically increases by $2\pi$, the orbitals $\varphi_n$ will evolve in to $\varphi_{n+1}$.
It is easy to see that $\mathcal{F}_x$ must proportional to the translation operator $\hat{T}_y$.
For $\kappa = \Phi_x$, the pieces of Eq.~\eqref{eq:U_masterEq} are
\begin{align}
	\theta^{(\varphi)}_n &= 2\pi\tau_x\frac{n}{N_\Phi}	,
&	A^{(\drehen)} &= -\tau_x\braket{\bondop{C}},
&	W_{ab} &= \delta_{a,b+1} e^{2\pi i\bondop{C}_{\bar{a}}}
\end{align}
($b+1$ refers to the MES which is $b$ translated by one site).
The flux matrix takes the form
\begin{align}
	\mathcal{F}_x &= \delta_{a,b+1} \mathcal{F}_y^{-\tau_x} \exp\left[ i\nu(-\Phi_y-\pi) \right]
	.	\label{eq:Fx}
\end{align}

\end{itemize}

In general, the flux matrices must satisfy the algebra
\begin{align}
	\mathcal{F}_x \mathcal{F}_y = e^{2\pi i\nu} \mathcal{F}_y \mathcal{F}_x	.
\end{align}

\subsection{\texorpdfstring
	{Dehn twist $\mathcal{T}$: topological spin, central charge, and Hall viscosity}
	{Dehn twist: topological spin, central charge, and Hall viscosity}}

Varying $\kappa = \tau_x$ from 0 to 1, we get the Berry phase $U_T$. Note that under $\mathcal{T}$, the fluxes transform as $\mathcal{T}: (\Phi_x, \Phi_y) \to (\Phi_x, \Phi_y + \Phi_x)$.
Thus we expect that at $\Phi_x=0$, $U_T$ should be unambiguously defined, while at $\Phi_x = \pi$, only $U_T^2$ should be unambiguously defined, a point we will return to later.

For the $l$\textsuperscript{th} Landau level, the individual parts of Eq.~\eqref{eq:U_masterEq} are
\begin{align}
	\theta^{(\varphi)}_n &= 2\pi\frac{n(n-N_\Phi)}{2N_\Phi} - (2l+1)\frac{L_x}{4L_y}	,
&	A^{(\drehen)} &= -\braket{\bondop{K}}	,
&	W_{ab} &= \delta_{ab} e^{2\pi i\bondop{K}_{\bar{a}}}	.
	\label{eq:T_terms}
\end{align}
(The lowest Landau level is labeled by $l=0$.) We remind the reader that for $\Phi_x \neq 0$, the site index $n$ and $\hat{K}_n$ include a fractional part $\frac{\Phi_x}{2 \pi}$, see Eq.~\ref{eq:cyl_orb}.
We note that the first term of $\theta^{(\varphi)}_n$ has been derived by Wen and Wang in Ref.~\onlinecite{WenWang-2008}.
Combining them, we have
\begin{align}
	U_{T;ab} &= \delta_{ab} \,
		\exp\left\{2\pi i \left[
			\bondop{K} - \dxp{\bondop{H}}
			+ \frac{\nu}{2} \left( \frac{\Phi_x^2}{4\pi^2} - \frac{\Phi_x}{2\pi} + \frac{1}{6} \right)
		 - \frac{(2l+1)\nu}{16 \pi^2 \ell_B^2} L_x^2 \right]  \right\}
	,	\label{eq:T}
\end{align}
where
\begin{align}
	\bondop{H}_{\bar{n}} = \bondop{K}_{\bar{n}} - \bar{n}\bondop{C}_{\bar{n}} + \bar{n}\dxp{\bondop{C}}	.
\end{align}
In light of Eq.~\eqref{eq:defP} (where $\dxp{\bondop{C}} = 0$), we can write $\bondop{H} = \bondop{P} + \frac{1}{2\nu}\bondop{C}^2$ 
	and interpret $\bondop{H}$ as the ``energy operator'' for auxiliary states.
$\bondop{H}$ is convenient as it is invariant under translation by $q$ sites.
As mentioned, at $\Phi_x=0$, $U_T$ is unambiguously defined because $\bondop{K} \bmod 1$ is single-valued on any bond (for a MES) and so Eq.~\eqref{eq:T} is well-defined.
For $\Phi_x = \pi$, however, only $\bondop{K} \bmod \frac12$ is single-valued which implies that only $U_T^2$ is well-defined (cf.~section on \hyperref[sec:Modular]{modular transformations}).
Finally, we also note that a term proportional to $N^e N_\Phi$ has been dropped from the result.

If we use a convention where $\dxp{\bondop{C}} = 0$, set $\Phi_x = \pi$ (which is most natural for fermions) and restrict to the lowest Landau level,
then Eq.~\eqref{eq:T} simplifies to
\begin{align}
	U_{T;ab} &=
		\exp\left\{2\pi i \left[
			\bondop{K} - \dxp{\bondop{K} - \bar{n}\bondop{C}_{\bar{n}}}
			- \frac{\nu}{24}
		 - \frac{\nu}{16 \pi^2 \ell_B^2} L_x^2 \right]  \right\}
	.
	&& \textrm{(When $l=0$, $\dxp{\bondop{C}} = 0$, and $\Phi_x=\pi$.)}
	\label{eq:T_piflux}
\end{align}
This is the form shown in the main text.

\textit{Comments.}
In both Eq.~\eqref{eq:T} and \eqref{eq:T_piflux}, the quantity $\bondop{K} - \dxp{\bondop{H}}$ measures the average momentum to the left of a cut, which is what distinguishes the various ground states from one another (allowing $h_a$ to be extracted).
There are terms dependent of $\nu$ because the formulas are written in terms of the orbital entanglement spectra, as opposed to the real-space entanglement.

In addition, in the thermodynamic limit $\braket{\bondop{C}_{\bar{n}}}$ and $\dxp{\bondop{C}}$ would be equal as long as charge-density wave order is absent.
(The same hold for $\bondop{H}$.)
Hence the formulas above are also useful is only the entanglement spectrum of a \emph{single cut} is known.

\textit{Discussion.}
	In order to interpret $U_T$, we picture the MES as a torus with a topological flux $a$ winding around the $y$-cycle.
We cut the torus at $y=0$, shear the segment of cylinder evenly throughout the bulk so that $y=0$ remains fixed while $y = L_y$ rotates by $L_x$, then reglue the ends.
Since an anyonic flux $a$ terminates at the edge, there is a quasiparticle $a$ on the edge $y = L_y$ which moves once around the circumference under the shear.
The quasiparticle has momentum $\frac{2 \pi}{L_x} ( h_a - c/24)$ (if the edge is not chiral, this should be understood as $h_a - \bar{h}_a - \frac{c - \bar{c}}{24}$),
	hence traversing a distance $L_x$ generates a phase $e^{2\pi i(h_a - c/24)} = \theta_a e^{-2\pi i c/24}$.
	($\theta_a$ is known as the `topological spin' of the quasiparticle $a$.)
The bulk itself shears as well, with the strain changing as $\frac{L_x}{L_y} d\tau_x$.
The finite `Hall viscosity' $\HallVis$ results in a phase per unit area proportional to the changing strain \cite{Avron-QHViscosity95},
\begin{align}
	\theta_\text{bulk}
	= \hbar^{-1} \int_{\tau_x=0}^1\!\! (L_xL_y) \HallVis \frac{L_x}{L_y} d\tau_x
	= \hbar^{-1} \HallVis L_x^2 	.
\end{align}	
Together, the expected result is	 	
\begin{align}
	U_{T;ab}
		&= \delta_{ab} \exp\left[ 2 \pi i \left( h_a - \frac{c}{24} - \frac{\HallVis}{2\pi\hbar} L_x^2 \right) \right]
	.	\label{eq:UT_hc}
\end{align}
The Hall viscosity is known to be related to the `shift' $\shift$ on a sphere via $\HallVis = \frac{\hbar}{4} \frac{\nu}{2 \pi \ell_B^2} \shift$ \cite{Read-QHViscosity09}.
	To extract the quantities independently, we can first fit the quadratic part to extract $\HallVis$ and isolate the constant part $h_a - c/24$.
Once the $\mathds{1}$ MES and bond is determined, we obtain $c$, and the remaining $h_a$'s can be read off directly from the ratio $e^{2\pi i h_a} = U_{T;aa} / U_{T;\mathds{1}\mathds{1}}$.

\begin{figure}[tb]
	\includegraphics[width=170mm]{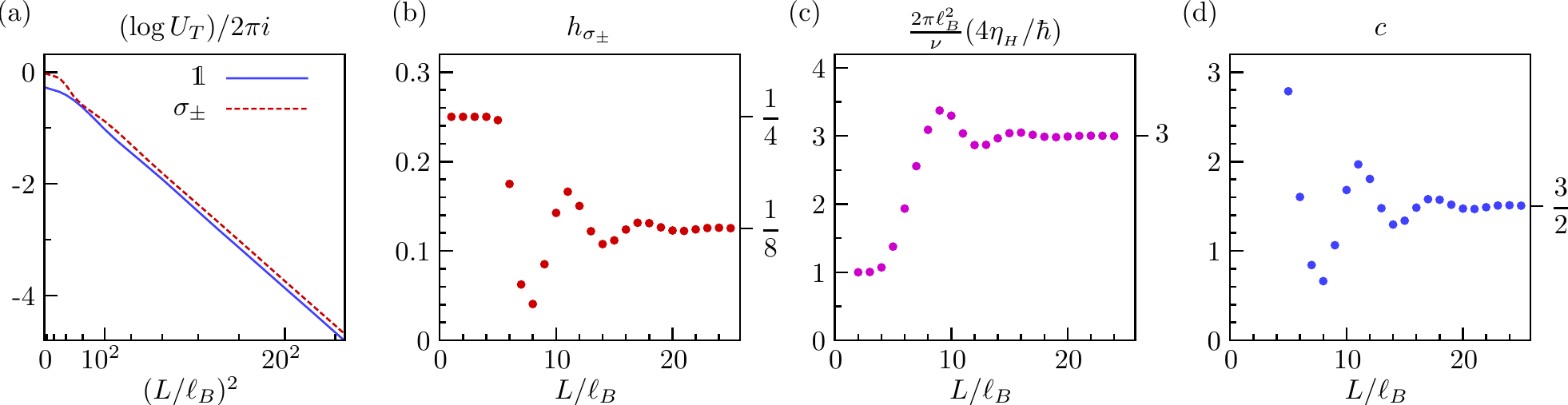}
	\caption{%
		Various quantities characterizing the topological Moore-Read phase at $\nu = 1/2$.
		The Berry phase $U_T$ arising from a Dehn twist is acquired from the model Moore-Read wave function via Eq.~\eqref{eq:T_piflux}
			and $h, c, \HallVis$ are extracted from Eq.~\eqref{eq:UT_hc} at various circumference $L = L_x$.
		(a) The argument of the phase $U_T$ plot vs.\ $L^2/\ell_B^2$.
			For large $L$, the argument becomes linear in $L^2$.
		(b) $h$ of the $\sigma_\pm$ quasiparticle, extracted from the ratio of $U_T$ between the $\sigma_\pm$ and $\mathds{1}$ ground states,
			via the ratio $\exp (2 \pi i h_{\sigma_\pm}) = U_T(\sigma_\pm) / U_T(\mathds{1})$.
		(c) The Hall viscosity $\HallVis$ extracted by fitting to the form $U_{T;\mathds{1}\mathds{1}} = e^{-2\pi i c/24} \exp(-\frac{\HallVis}{2\pi\hbar}L^2)$ for the identity sector.
			The data is presented as the `shift' $\shift = (2\pi\ell_B^2 / \nu) (4\HallVis / \hbar)$.
		(d) The chiral central charge extracted from $U_{T;\mathds{1}\mathds{1}}$, assuming $\shift = 3$ for the Moore-Read state.
		(b)-(d) In all cases $L/\ell_B$ must be sufficiently large for these topological quantities to be reliably extracted from the entanglement spectrum.
	}
	\label{fig:mr_dehn}
\end{figure}		

	In Ref.~\onlinecite{WenWang-2008}, the differences $h_a - h_b$ can be extracted from `pattern of zeros,' which can be understood as the $L\to0$ limit in which the state becomes a $\chi=1$ tensor product.
In this limit, only the term $\theta^{(\varphi)}_n$ from Eq.~\eqref{eq:T_terms} contributes to the Berry phase $U_T$ and hence 
	the differences $h_a - h_b$ inferred from Eq.~\ref{eq:T} reproduces Wen and Wang's result (Eq.~(32) and (33) of Ref.~\cite{WenWang-2008}).
However, in the limit $L \to \infty$, the ratio $\theta_a/\theta_b = e^{2\pi i(h_a - h_b)}$ computed via Eq.~\eqref{eq:T} matches that of Ref.~\onlinecite{WenWang-2008} \emph{only} when the quasiparticles $a$, $b$ are related by attaching fractional fluxes (in the MES language, the states are related by translation).
Hence in the (single-Landau level) Abelian case, where all MES are related by flux attachment, the spins $h_a$'s can be recovered in the limit $L \to 0$ (using $h_{\mathds{1}} = 0$).
But for non-Abelian cases, where not all MES are related by translation, only in the $L \to \infty$ limit gives the correct result,
	so the result of Ref.~\onlinecite{WenWang-2008} [Eq.~(33)] appears to be valid only for Abelian phases.

As an example, consider the Moore-Read state at $\nu=\frac12$.
The MR state contains two non-Abelian quasiparticle `$\sigma_+$' and `$\sigma_-$',
	with charges $\pm \frac{e}{4}$ and both with quantum dimension $d_\sigma = \sqrt{2}$.
Its topological spin is expected to be $h_{\sigma_\pm} = \frac{1}{2\nu}Q^2 + h_\sigma = \frac{1}{16} + \frac{1}{16} = \frac18$.
The edge of the MR phase consists of a free boson and Majorana mode with combined chiral central charge of $1 + \frac12 = \frac32$.
In Fig.~\ref{fig:mr_dehn}, we plot the values of $h$, $c$ and $\HallVis$ extracted by fitting $U_T$ of the model MR wavefunctions \cite{MooreRead1991} to Eq.~\eqref{eq:UT_hc} at various $L_x$.

\subsection{Modular transformations}
\label{sec:Modular}

The modular transformations are affine maps from the torus to itself, the set of which is the modular group
	$\mathrm{PSL}(2, \mathbb{Z}) \cong \mathrm{SL}(2, \mathbb{Z}) / \mathbb{Z}_2$.
For example, the `$T$' transformation corresponds to a Dehn twist sending $\tau \rightarrow \tau+1$ (the same as $\tau_x \rightarrow \tau_x+1$).
The `$S$' transformation rotates the torus sending $\tau \rightarrow -1/\tau$.
	(When $\tau_x = 0$, this corresponds to a $\pi/2$ rotation swapping $L_x$ with $L_y$.)
Since $T$ and $S$ generate the entire modular group, we focus only on these two transformations.

The $\mathcal{T}$- and $\mathcal{S}$-matrices describes how the set of ground states transform under their respective modular transformations \cite{Keski-Vakkuri-1993}.
As discussed in the previous section, $\mathcal{T}$ is a diagonal matrix with entries $\theta_a$ known as the `topological spin', the action of rotating a quasiparticle type $a$ by $2\pi$.
$\mathcal{S}_{ab}$ gives the mutual statistics of braiding $a$ and $b$ around each other.
Generically $\mathcal{T}, \mathcal{S}$ are elements in a projective representation of $\mathrm{SL}(2, \mathbb{Z})$; the double cover of the modular group.
The double cover is necessary because $S^2 = 1 \in \mathrm{PSL}(2, \mathbb{Z})$, but $\mathcal{S}^2$ corresponds to charge-conjugation and is not a multiple of the identity; rather $\mathcal{S}^4 \propto \mathds{1}$.

Note that under the modular transformations the fluxes transform as $\mathcal{T}: (\Phi_x, \Phi_y) \to (\Phi_x, \Phi_y + \Phi_x)$, $\mathcal{S}: (\Phi_x, \Phi_y) \to (\Phi_y, -\Phi_x)$.
As we have discussed, $\Phi_i = 0$ is most natural for bosons, but $\Phi_i = \pi$ is most natural for fermions, so we must return to this subtlety.

\begin{itemize}
\item{Constraining (or determining) $\mathcal{S}$ from $\mathcal{F}$.}
When $\tau_x = 0$, we can use $\mathcal{S}$ to relate the two flux matrices \cite{WenWang-2008},
\begin{align}
	\mathcal{F}_y &= \mathcal{S} \mathcal{F}_x \mathcal{S}^{-1}	.
	\label{eq:S_from_F}
\end{align}
While this alone cannot be used to solve for $\mathcal{S}$, there are additional constraints,
\begin{align}\label{eq:S_constraint}
		\mathcal{S}_{\mathds{1}a} = \frac{d_a}{\mathcal{D}} = e^{-\gamma_a}	, \quad \mathcal{S}_{ab} = \mathcal{S}_{ba}	,
\end{align}
where $d_a$ is the quantum dimension for the quasiparticle $a$, with $\mathcal{D} = \sqrt{\sum_a d_a^2}$ being the total quantum dimension.
$\gamma_a$ is the topological entanglement entropy for the MES $\ket{\Xi_a}$ defined in the main text.
For certain phases such as the Moore-Read state and the $\nu=2/5$ Jain state the modular $\mathcal{S}$-matrix may be determined from these constraints alone.
Solving for $\mathcal{S}$ in the MES basis essentially amounts to diagonalizing $\mathcal{F}_x$.


\item{Flux sectors and modular transformations.}
A subtlety in the computing $\mathcal{T}$ and $\mathcal{S}$ is the interplay of modular transformations with boundary conditions ($\Phi_x, \Phi_y$).
Since $T, S$ change the fluxes, we need to instead consider a larger Hilbert space for which the boundary conditions may take on four possible combinations:
	$(\Phi_x,\Phi_y) \in \{ (0,0), (0,\pi), (\pi,0), (\pi,\pi)\}$, which as a shorthand we refer to as \textbf{PP}, \textbf{PA}, \textbf{AP}, \textbf{AA}, respectively.
Each of these sectors consist of $\gsd$ linear independent ground states, for a total of $4\gsd$-dimensional ground state manifold.
The \textbf{PP} sector is closed under the action of $\mathcal{T}$ and $\mathcal{S}$, but the other three sectors mixes under modular transformations \cite{Ginsparg}.

We write $\mathcal{T}$ and $\mathcal{S}$ as block matrices, where each block is an $\gsd\times\gsd$ matrix describing transitions between sectors.
The order of the four columns/rows are \textbf{PP}, \textbf{PA}, \textbf{AP}, \textbf{AA}.
\begin{align}
	\mathcal{T} &=
		\left[\!\begin{array}{c|c|c|c}
			\mathcal{T}^\textbf{PP}	&&&	\\\hline
			&	\mathcal{T}^\textbf{PA}	&&	\\\hline
			&&&	\mathcal{T}^+	\\\hline
			&&	\mathcal{T}^-	&
		\end{array}\!\right]	,
&	\mathcal{S} &=
		\left[\!\begin{array}{c|c|c|c}
			\mathcal{S}^\textbf{PP}	&&&	\\\hline
			&&	\mathcal{S}^+	&	\\\hline
			&	\mathcal{S}^-	&&	\\\hline
			&&&	\mathcal{S}^\textbf{AA}
		\end{array}\!\right]	.
\end{align}
In the minimal entangled basis, each $\mathcal{T}$-submatrix are still diagonal.
In the \textbf{AP} and \textbf{AA} sectors ($\Phi_x=\pi$), the formula Eq.~\eqref{eq:T} \emph{squared} gives the product $\mathcal{T}^+\mathcal{T}^-$, as two Dehn twists are required to come back to the same wavefunction.
In other words, Eq.~\eqref{eq:T} will only give $h$ modulo $\frac12$.

\end{itemize}

\end{widetext}

\ifdefined\qhsubputreferences \bibliography{\qhsubputreferences} \else {} \fi

%% file: qh_idmrg.bbl
\begin{thebibliography}{60}%
\makeatletter
\providecommand \@ifxundefined [1]{%
 \@ifx{#1\undefined}
}%
\providecommand \@ifnum [1]{%
 \ifnum #1\expandafter \@firstoftwo
 \else \expandafter \@secondoftwo
 \fi
}%
\providecommand \@ifx [1]{%
 \ifx #1\expandafter \@firstoftwo
 \else \expandafter \@secondoftwo
 \fi
}%
\providecommand \natexlab [1]{#1}%
\providecommand \enquote  [1]{``#1''}%
\providecommand \bibnamefont  [1]{#1}%
\providecommand \bibfnamefont [1]{#1}%
\providecommand \citenamefont [1]{#1}%
\providecommand \href@noop [0]{\@secondoftwo}%
\providecommand \href [0]{\begingroup \@sanitize@url \@href}%
\providecommand \@href[1]{\@@startlink{#1}\@@href}%
\providecommand \@@href[1]{\endgroup#1\@@endlink}%
\providecommand \@sanitize@url [0]{\catcode `\\12\catcode `\$12\catcode
  `\&12\catcode `\#12\catcode `\^12\catcode `\_12\catcode `\%12\relax}%
\providecommand \@@startlink[1]{}%
\providecommand \@@endlink[0]{}%
\providecommand \url  [0]{\begingroup\@sanitize@url \@url }%
\providecommand \@url [1]{\endgroup\@href {#1}{\urlprefix }}%
\providecommand \urlprefix  [0]{URL }%
\providecommand \Eprint [0]{\href }%
\providecommand \doibase [0]{http://dx.doi.org/}%
\providecommand \selectlanguage [0]{\@gobble}%
\providecommand \bibinfo  [0]{\@secondoftwo}%
\providecommand \bibfield  [0]{\@secondoftwo}%
\providecommand \translation [1]{[#1]}%
\providecommand \BibitemOpen [0]{}%
\providecommand \bibitemStop [0]{}%
\providecommand \bibitemNoStop [0]{.\EOS\space}%
\providecommand \EOS [0]{\spacefactor3000\relax}%
\providecommand \BibitemShut  [1]{\csname bibitem#1\endcsname}%
\let\auto@bib@innerbib\@empty
\bibitem [{\citenamefont {Wen}(1990)}]{Wen-1990}%
  \BibitemOpen
  \bibfield  {author} {\bibinfo {author} {\bibfnamefont {X.-G.}\ \bibnamefont
  {Wen}},\ }\href@noop {} {\bibfield  {journal} {\bibinfo  {journal} {Int. J.
  Mod. Phys.}\ }\textbf {\bibinfo {volume} {B4}},\ \bibinfo {pages} {239}
  (\bibinfo {year} {1990})}\BibitemShut {NoStop}%
\bibitem [{\citenamefont {Wen}\ and\ \citenamefont {Niu}(1990)}]{WenNiu-1990}%
  \BibitemOpen
  \bibfield  {author} {\bibinfo {author} {\bibfnamefont {X.~G.}\ \bibnamefont
  {Wen}}\ and\ \bibinfo {author} {\bibfnamefont {Q.}~\bibnamefont {Niu}},\
  }\href {\doibase 10.1103/PhysRevB.41.9377} {\bibfield  {journal} {\bibinfo
  {journal} {Phys. Rev. B}\ }\textbf {\bibinfo {volume} {41}},\ \bibinfo
  {pages} {9377} (\bibinfo {year} {1990})}\BibitemShut {NoStop}%
\bibitem [{\citenamefont {Tsui}\ \emph {et~al.}(1982)\citenamefont {Tsui},
  \citenamefont {Stormer},\ and\ \citenamefont {Gossard}}]{Tsui-1982}%
  \BibitemOpen
  \bibfield  {author} {\bibinfo {author} {\bibfnamefont {D.~C.}\ \bibnamefont
  {Tsui}}, \bibinfo {author} {\bibfnamefont {H.~L.}\ \bibnamefont {Stormer}}, \
  and\ \bibinfo {author} {\bibfnamefont {A.~C.}\ \bibnamefont {Gossard}},\
  }\href {\doibase 10.1103/PhysRevLett.48.1559} {\bibfield  {journal} {\bibinfo
   {journal} {Phys. Rev. Lett.}\ }\textbf {\bibinfo {volume} {48}},\ \bibinfo
  {pages} {1559} (\bibinfo {year} {1982})}\BibitemShut {NoStop}%
\bibitem [{\citenamefont {Kitaev}(2003)}]{Kitaev-1997}%
  \BibitemOpen
  \bibfield  {author} {\bibinfo {author} {\bibfnamefont {A.~Y.}\ \bibnamefont
  {Kitaev}},\ }\href {\doibase 10.1016/S0003-4916(02)00018-0} {\bibfield
  {journal} {\bibinfo  {journal} {Ann. Phys.}\ }\textbf {\bibinfo {volume}
  {303}},\ \bibinfo {pages} {2} (\bibinfo {year} {2003})}\BibitemShut {NoStop}%
\bibitem [{\citenamefont {Das~Sarma}\ \emph {et~al.}(2005)\citenamefont
  {Das~Sarma}, \citenamefont {Freedman},\ and\ \citenamefont
  {Nayak}}]{DasSarmaInterferometer-2005}%
  \BibitemOpen
  \bibfield  {author} {\bibinfo {author} {\bibfnamefont {S.}~\bibnamefont
  {Das~Sarma}}, \bibinfo {author} {\bibfnamefont {M.}~\bibnamefont {Freedman}},
  \ and\ \bibinfo {author} {\bibfnamefont {C.}~\bibnamefont {Nayak}},\ }\href
  {\doibase 10.1103/PhysRevLett.94.166802} {\bibfield  {journal} {\bibinfo
  {journal} {Phys. Rev. Lett.}\ }\textbf {\bibinfo {volume} {94}},\ \bibinfo
  {pages} {166802} (\bibinfo {year} {2005})}\BibitemShut {NoStop}%
\bibitem [{\citenamefont {Bonderson}\ \emph {et~al.}(2006)\citenamefont
  {Bonderson}, \citenamefont {Kitaev},\ and\ \citenamefont
  {Shtengel}}]{BondersonInterferometer-2006}%
  \BibitemOpen
  \bibfield  {author} {\bibinfo {author} {\bibfnamefont {P.}~\bibnamefont
  {Bonderson}}, \bibinfo {author} {\bibfnamefont {A.}~\bibnamefont {Kitaev}}, \
  and\ \bibinfo {author} {\bibfnamefont {K.}~\bibnamefont {Shtengel}},\ }\href
  {\doibase 10.1103/PhysRevLett.96.016803} {\bibfield  {journal} {\bibinfo
  {journal} {Phys. Rev. Lett.}\ }\textbf {\bibinfo {volume} {96}},\ \bibinfo
  {pages} {016803} (\bibinfo {year} {2006})}\BibitemShut {NoStop}%
\bibitem [{\citenamefont {Nayak}\ \emph {et~al.}(2008)\citenamefont {Nayak},
  \citenamefont {Simon}, \citenamefont {Stern}, \citenamefont {Freedman},\ and\
  \citenamefont {Das~Sarma}}]{Nayak-2008}%
  \BibitemOpen
  \bibfield  {author} {\bibinfo {author} {\bibfnamefont {C.}~\bibnamefont
  {Nayak}}, \bibinfo {author} {\bibfnamefont {S.~H.}\ \bibnamefont {Simon}},
  \bibinfo {author} {\bibfnamefont {A.}~\bibnamefont {Stern}}, \bibinfo
  {author} {\bibfnamefont {M.}~\bibnamefont {Freedman}}, \ and\ \bibinfo
  {author} {\bibfnamefont {S.}~\bibnamefont {Das~Sarma}},\ }\href {\doibase
  10.1103/RevModPhys.80.1083} {\bibfield  {journal} {\bibinfo  {journal} {Rev.
  Mod. Phys.}\ }\textbf {\bibinfo {volume} {80}},\ \bibinfo {eid} {1083}
  (\bibinfo {year} {2008})}\BibitemShut {NoStop}%
\bibitem [{\citenamefont {Laughlin}(1983)}]{Laughlin-1983}%
  \BibitemOpen
  \bibfield  {author} {\bibinfo {author} {\bibfnamefont {R.~B.}\ \bibnamefont
  {Laughlin}},\ }\href {\doibase 10.1103/PhysRevLett.50.1395} {\bibfield
  {journal} {\bibinfo  {journal} {Phys. Rev. Lett.}\ }\textbf {\bibinfo
  {volume} {50}},\ \bibinfo {pages} {1395} (\bibinfo {year}
  {1983})}\BibitemShut {NoStop}%
\bibitem [{\citenamefont {Yoshioka}\ \emph {et~al.}(1983)\citenamefont
  {Yoshioka}, \citenamefont {Halperin},\ and\ \citenamefont
  {Lee}}]{Yoshioka-1983}%
  \BibitemOpen
  \bibfield  {author} {\bibinfo {author} {\bibfnamefont {D.}~\bibnamefont
  {Yoshioka}}, \bibinfo {author} {\bibfnamefont {B.~I.}\ \bibnamefont
  {Halperin}}, \ and\ \bibinfo {author} {\bibfnamefont {P.~A.}\ \bibnamefont
  {Lee}},\ }\href {\doibase 10.1103/PhysRevLett.50.1219} {\bibfield  {journal}
  {\bibinfo  {journal} {Phys. Rev. Lett.}\ }\textbf {\bibinfo {volume} {50}},\
  \bibinfo {pages} {1219} (\bibinfo {year} {1983})}\BibitemShut {NoStop}%
\bibitem [{\citenamefont {Haldane}(1983)}]{Haldane-1983}%
  \BibitemOpen
  \bibfield  {author} {\bibinfo {author} {\bibfnamefont {F.~D.~M.}\
  \bibnamefont {Haldane}},\ }\href {\doibase 10.1103/PhysRevLett.51.605}
  {\bibfield  {journal} {\bibinfo  {journal} {Phys. Rev. Lett.}\ }\textbf
  {\bibinfo {volume} {51}},\ \bibinfo {pages} {605} (\bibinfo {year}
  {1983})}\BibitemShut {NoStop}%
\bibitem [{\citenamefont {Kitaev}\ and\ \citenamefont
  {Preskill}(2006)}]{KitaevPreskill}%
  \BibitemOpen
  \bibfield  {author} {\bibinfo {author} {\bibfnamefont {A.}~\bibnamefont
  {Kitaev}}\ and\ \bibinfo {author} {\bibfnamefont {J.}~\bibnamefont
  {Preskill}},\ }\href {\doibase 10.1103/PhysRevLett.96.110404} {\bibfield
  {journal} {\bibinfo  {journal} {Phys. Rev. Lett.}\ }\textbf {\bibinfo
  {volume} {96}},\ \bibinfo {eid} {110404} (\bibinfo {year}
  {2006})}\BibitemShut {NoStop}%
\bibitem [{\citenamefont {Levin}\ and\ \citenamefont {Wen}(2006)}]{Levin-2006}%
  \BibitemOpen
  \bibfield  {author} {\bibinfo {author} {\bibfnamefont {M.}~\bibnamefont
  {Levin}}\ and\ \bibinfo {author} {\bibfnamefont {X.-G.}\ \bibnamefont
  {Wen}},\ }\href {\doibase 10.1103/PhysRevLett.96.110405} {\bibfield
  {journal} {\bibinfo  {journal} {Phys. Rev. Lett.}\ }\textbf {\bibinfo
  {volume} {96}},\ \bibinfo {eid} {110405} (\bibinfo {year}
  {2006})}\BibitemShut {NoStop}%
\bibitem [{\citenamefont {Li}\ and\ \citenamefont {Haldane}(2008)}]{Li-2008}%
  \BibitemOpen
  \bibfield  {author} {\bibinfo {author} {\bibfnamefont {H.}~\bibnamefont
  {Li}}\ and\ \bibinfo {author} {\bibfnamefont {F.~D.~M.}\ \bibnamefont
  {Haldane}},\ }\href {\doibase 10.1103/PhysRevLett.101.010504} {\bibfield
  {journal} {\bibinfo  {journal} {Phys. Rev. Lett.}\ }\textbf {\bibinfo
  {volume} {101}},\ \bibinfo {pages} {010504} (\bibinfo {year}
  {2008})}\BibitemShut {NoStop}%
\bibitem [{\citenamefont {Laeuchli}\ \emph {et~al.}(2010)\citenamefont
  {Laeuchli}, \citenamefont {Bergholtz},\ and\ \citenamefont
  {Haque}}]{Laeuchli-2010}%
  \BibitemOpen
  \bibfield  {author} {\bibinfo {author} {\bibfnamefont {A.~M.}\ \bibnamefont
  {Laeuchli}}, \bibinfo {author} {\bibfnamefont {E.~J.}\ \bibnamefont
  {Bergholtz}}, \ and\ \bibinfo {author} {\bibfnamefont {M.}~\bibnamefont
  {Haque}},\ }\href@noop {} {\bibfield  {journal} {\bibinfo  {journal} {New J.
  Phys.}\ }\textbf {\bibinfo {volume} {12}},\ \bibinfo {pages} {075004}
  (\bibinfo {year} {2010})}\BibitemShut {NoStop}%
\bibitem [{\citenamefont {Papi\ifmmode~\acute{c}\else \'{c}\fi{}}\ \emph
  {et~al.}(2011)\citenamefont {Papi\ifmmode~\acute{c}\else \'{c}\fi{}},
  \citenamefont {Bernevig},\ and\ \citenamefont {Regnault}}]{Papic-2011}%
  \BibitemOpen
  \bibfield  {author} {\bibinfo {author} {\bibfnamefont {Z.}~\bibnamefont
  {Papi\ifmmode~\acute{c}\else \'{c}\fi{}}}, \bibinfo {author} {\bibfnamefont
  {B.~A.}\ \bibnamefont {Bernevig}}, \ and\ \bibinfo {author} {\bibfnamefont
  {N.}~\bibnamefont {Regnault}},\ }\href {\doibase
  10.1103/PhysRevLett.106.056801} {\bibfield  {journal} {\bibinfo  {journal}
  {Phys. Rev. Lett.}\ }\textbf {\bibinfo {volume} {106}},\ \bibinfo {pages}
  {056801} (\bibinfo {year} {2011})}\BibitemShut {NoStop}%
\bibitem [{\citenamefont {Zhang}\ \emph {et~al.}(2012)\citenamefont {Zhang},
  \citenamefont {Grover}, \citenamefont {Turner}, \citenamefont {Oshikawa},\
  and\ \citenamefont {Vishwanath}}]{Zhang-2012}%
  \BibitemOpen
  \bibfield  {author} {\bibinfo {author} {\bibfnamefont {Y.}~\bibnamefont
  {Zhang}}, \bibinfo {author} {\bibfnamefont {T.}~\bibnamefont {Grover}},
  \bibinfo {author} {\bibfnamefont {A.}~\bibnamefont {Turner}}, \bibinfo
  {author} {\bibfnamefont {M.}~\bibnamefont {Oshikawa}}, \ and\ \bibinfo
  {author} {\bibfnamefont {A.}~\bibnamefont {Vishwanath}},\ }\href {\doibase
  10.1103/PhysRevB.85.235151} {\bibfield  {journal} {\bibinfo  {journal} {Phys.
  Rev. B}\ }\textbf {\bibinfo {volume} {85}},\ \bibinfo {pages} {235151}
  (\bibinfo {year} {2012})}\BibitemShut {NoStop}%
\bibitem [{\citenamefont {Zaletel}\ and\ \citenamefont
  {Mong}(2012)}]{Zaletel-2012}%
  \BibitemOpen
  \bibfield  {author} {\bibinfo {author} {\bibfnamefont {M.~P.}\ \bibnamefont
  {Zaletel}}\ and\ \bibinfo {author} {\bibfnamefont {R.~S.~K.}\ \bibnamefont
  {Mong}},\ }\href {\doibase 10.1103/PhysRevB.86.245305} {\bibfield  {journal}
  {\bibinfo  {journal} {Phys. Rev. B}\ }\textbf {\bibinfo {volume} {86}},\
  \bibinfo {pages} {245305} (\bibinfo {year} {2012})}\BibitemShut {NoStop}%
\bibitem [{\citenamefont {Estienne}\ \emph {et~al.}(2013)\citenamefont
  {Estienne}, \citenamefont {Papi\ifmmode~\acute{c}\else \'{c}\fi{}},
  \citenamefont {Regnault},\ and\ \citenamefont {Bernevig}}]{estienne-2012}%
  \BibitemOpen
  \bibfield  {author} {\bibinfo {author} {\bibfnamefont {B.}~\bibnamefont
  {Estienne}}, \bibinfo {author} {\bibfnamefont {Z.}~\bibnamefont
  {Papi\ifmmode~\acute{c}\else \'{c}\fi{}}}, \bibinfo {author} {\bibfnamefont
  {N.}~\bibnamefont {Regnault}}, \ and\ \bibinfo {author} {\bibfnamefont
  {B.~A.}\ \bibnamefont {Bernevig}},\ }\href {\doibase
  10.1103/PhysRevB.87.161112} {\bibfield  {journal} {\bibinfo  {journal} {Phys.
  Rev. B}\ }\textbf {\bibinfo {volume} {87}},\ \bibinfo {pages} {161112}
  (\bibinfo {year} {2013})}\BibitemShut {NoStop}%
\bibitem [{\citenamefont {White}(1992)}]{White-1992}%
  \BibitemOpen
  \bibfield  {author} {\bibinfo {author} {\bibfnamefont {S.~R.}\ \bibnamefont
  {White}},\ }\href {\doibase 10.1103/PhysRevLett.69.2863} {\bibfield
  {journal} {\bibinfo  {journal} {Phys. Rev. Lett.}\ }\textbf {\bibinfo
  {volume} {69}},\ \bibinfo {pages} {2863} (\bibinfo {year}
  {1992})}\BibitemShut {NoStop}%
\bibitem [{\citenamefont {Shibata}\ and\ \citenamefont
  {Yoshioka}(2001)}]{Naokazu-2001}%
  \BibitemOpen
  \bibfield  {author} {\bibinfo {author} {\bibfnamefont {N.}~\bibnamefont
  {Shibata}}\ and\ \bibinfo {author} {\bibfnamefont {D.}~\bibnamefont
  {Yoshioka}},\ }\href {\doibase 10.1103/PhysRevLett.86.5755} {\bibfield
  {journal} {\bibinfo  {journal} {Phys. Rev. Lett.}\ }\textbf {\bibinfo
  {volume} {86}},\ \bibinfo {pages} {5755} (\bibinfo {year}
  {2001})}\BibitemShut {NoStop}%
\bibitem [{\citenamefont {Bergholtz}\ and\ \citenamefont
  {Karlhede}(2003)}]{BergholtzKarlhede-2003}%
  \BibitemOpen
  \bibfield  {author} {\bibinfo {author} {\bibfnamefont {E.~J.}\ \bibnamefont
  {Bergholtz}}\ and\ \bibinfo {author} {\bibfnamefont {A.}~\bibnamefont
  {Karlhede}},\ }\href@noop {} {\enquote {\bibinfo {title} {{Density Matrix
  Renormalization Group Study of a Lowest Landau Level Electron Gas on a Thin
  Cylinder}},}\ } (\bibinfo {year} {2003}),\ \bibinfo {note} {unpublished},\
  \Eprint {http://arxiv.org/abs/cond-mat/0304517} {arXiv:cond-mat/0304517}
  \BibitemShut {NoStop}%
\bibitem [{\citenamefont {Feiguin}\ \emph {et~al.}(2008)\citenamefont
  {Feiguin}, \citenamefont {Rezayi}, \citenamefont {Nayak},\ and\ \citenamefont
  {Das~Sarma}}]{Feiguin-2008}%
  \BibitemOpen
  \bibfield  {author} {\bibinfo {author} {\bibfnamefont {A.~E.}\ \bibnamefont
  {Feiguin}}, \bibinfo {author} {\bibfnamefont {E.}~\bibnamefont {Rezayi}},
  \bibinfo {author} {\bibfnamefont {C.}~\bibnamefont {Nayak}}, \ and\ \bibinfo
  {author} {\bibfnamefont {S.}~\bibnamefont {Das~Sarma}},\ }\href {\doibase
  10.1103/PhysRevLett.100.166803} {\bibfield  {journal} {\bibinfo  {journal}
  {Phys. Rev. Lett.}\ }\textbf {\bibinfo {volume} {100}},\ \bibinfo {pages}
  {166803} (\bibinfo {year} {2008})}\BibitemShut {NoStop}%
\bibitem [{\citenamefont {Kovrizhin}(2010)}]{Kovrizhin-2010}%
  \BibitemOpen
  \bibfield  {author} {\bibinfo {author} {\bibfnamefont {D.~L.}\ \bibnamefont
  {Kovrizhin}},\ }\href {\doibase 10.1103/PhysRevB.81.125130} {\bibfield
  {journal} {\bibinfo  {journal} {Phys. Rev. B}\ }\textbf {\bibinfo {volume}
  {81}},\ \bibinfo {pages} {125130} (\bibinfo {year} {2010})}\BibitemShut
  {NoStop}%
\bibitem [{\citenamefont {Zhao}\ \emph {et~al.}(2011)\citenamefont {Zhao},
  \citenamefont {Sheng},\ and\ \citenamefont {Haldane}}]{Zhao-2011}%
  \BibitemOpen
  \bibfield  {author} {\bibinfo {author} {\bibfnamefont {J.}~\bibnamefont
  {Zhao}}, \bibinfo {author} {\bibfnamefont {D.~N.}\ \bibnamefont {Sheng}}, \
  and\ \bibinfo {author} {\bibfnamefont {F.~D.~M.}\ \bibnamefont {Haldane}},\
  }\href {\doibase 10.1103/PhysRevB.83.195135} {\bibfield  {journal} {\bibinfo
  {journal} {Phys. Rev. B}\ }\textbf {\bibinfo {volume} {83}},\ \bibinfo
  {pages} {195135} (\bibinfo {year} {2011})}\BibitemShut {NoStop}%
\bibitem [{\citenamefont {Hu}\ \emph {et~al.}(2012)\citenamefont {Hu},
  \citenamefont {Papic}, \citenamefont {Johri}, \citenamefont {Bhatt},\ and\
  \citenamefont {Schmitteckert}}]{Hu-2012}%
  \BibitemOpen
  \bibfield  {author} {\bibinfo {author} {\bibfnamefont {Z.-X.}\ \bibnamefont
  {Hu}}, \bibinfo {author} {\bibfnamefont {Z.}~\bibnamefont {Papic}}, \bibinfo
  {author} {\bibfnamefont {S.}~\bibnamefont {Johri}}, \bibinfo {author}
  {\bibfnamefont {R.}~\bibnamefont {Bhatt}}, \ and\ \bibinfo {author}
  {\bibfnamefont {P.}~\bibnamefont {Schmitteckert}},\ }\href {\doibase
  10.1016/j.physleta.2012.05.031} {\bibfield  {journal} {\bibinfo  {journal}
  {Physics Letters A}\ }\textbf {\bibinfo {volume} {376}},\ \bibinfo {pages}
  {2157 } (\bibinfo {year} {2012})}\BibitemShut {NoStop}%
\bibitem [{\citenamefont {L\"auchli}\ \emph {et~al.}(2010)\citenamefont
  {L\"auchli}, \citenamefont {Bergholtz}, \citenamefont {Suorsa},\ and\
  \citenamefont {Haque}}]{LauchliBergholtz-2010}%
  \BibitemOpen
  \bibfield  {author} {\bibinfo {author} {\bibfnamefont {A.~M.}\ \bibnamefont
  {L\"auchli}}, \bibinfo {author} {\bibfnamefont {E.~J.}\ \bibnamefont
  {Bergholtz}}, \bibinfo {author} {\bibfnamefont {J.}~\bibnamefont {Suorsa}}, \
  and\ \bibinfo {author} {\bibfnamefont {M.}~\bibnamefont {Haque}},\ }\href
  {\doibase 10.1103/PhysRevLett.104.156404} {\bibfield  {journal} {\bibinfo
  {journal} {Phys. Rev. Lett.}\ }\textbf {\bibinfo {volume} {104}},\ \bibinfo
  {pages} {156404} (\bibinfo {year} {2010})}\BibitemShut {NoStop}%
\bibitem [{\citenamefont {Liu}\ \emph {et~al.}(2012)\citenamefont {Liu},
  \citenamefont {Bergholtz}, \citenamefont {Fan},\ and\ \citenamefont
  {L\"auchli}}]{ZhaoBergholtz2012}%
  \BibitemOpen
  \bibfield  {author} {\bibinfo {author} {\bibfnamefont {Z.}~\bibnamefont
  {Liu}}, \bibinfo {author} {\bibfnamefont {E.~J.}\ \bibnamefont {Bergholtz}},
  \bibinfo {author} {\bibfnamefont {H.}~\bibnamefont {Fan}}, \ and\ \bibinfo
  {author} {\bibfnamefont {A.~M.}\ \bibnamefont {L\"auchli}},\ }\href {\doibase
  10.1103/PhysRevB.85.045119} {\bibfield  {journal} {\bibinfo  {journal} {Phys.
  Rev. B}\ }\textbf {\bibinfo {volume} {85}},\ \bibinfo {pages} {045119}
  (\bibinfo {year} {2012})}\BibitemShut {NoStop}%
\bibitem [{\citenamefont {Dong}\ \emph {et~al.}(2008)\citenamefont {Dong},
  \citenamefont {Fradkin}, \citenamefont {Leigh},\ and\ \citenamefont
  {Nowling}}]{Dong-2008}%
  \BibitemOpen
  \bibfield  {author} {\bibinfo {author} {\bibfnamefont {S.}~\bibnamefont
  {Dong}}, \bibinfo {author} {\bibfnamefont {E.}~\bibnamefont {Fradkin}},
  \bibinfo {author} {\bibfnamefont {R.~G.}\ \bibnamefont {Leigh}}, \ and\
  \bibinfo {author} {\bibfnamefont {S.}~\bibnamefont {Nowling}},\ }\href@noop
  {} {\bibfield  {journal} {\bibinfo  {journal} {Journal of High Energy
  Physics}\ }\textbf {\bibinfo {volume} {2008}},\ \bibinfo {pages} {016}
  (\bibinfo {year} {2008})}\BibitemShut {NoStop}%
\bibitem [{\citenamefont {Avron}\ \emph {et~al.}(1995)\citenamefont {Avron},
  \citenamefont {Seiler},\ and\ \citenamefont {Zograf}}]{Avron-QHViscosity95}%
  \BibitemOpen
  \bibfield  {author} {\bibinfo {author} {\bibfnamefont {J.~E.}\ \bibnamefont
  {Avron}}, \bibinfo {author} {\bibfnamefont {R.}~\bibnamefont {Seiler}}, \
  and\ \bibinfo {author} {\bibfnamefont {P.~G.}\ \bibnamefont {Zograf}},\
  }\href {\doibase 10.1103/PhysRevLett.75.697} {\bibfield  {journal} {\bibinfo
  {journal} {Phys. Rev. Lett.}\ }\textbf {\bibinfo {volume} {75}},\ \bibinfo
  {pages} {697} (\bibinfo {year} {1995})}\BibitemShut {NoStop}%
\bibitem [{\citenamefont {Read}(2009)}]{Read-QHViscosity09}%
  \BibitemOpen
  \bibfield  {author} {\bibinfo {author} {\bibfnamefont {N.}~\bibnamefont
  {Read}},\ }\href {\doibase 10.1103/PhysRevB.79.045308} {\bibfield  {journal}
  {\bibinfo  {journal} {Phys. Rev. B}\ }\textbf {\bibinfo {volume} {79}},\
  \bibinfo {pages} {045308} (\bibinfo {year} {2009})}\BibitemShut {NoStop}%
\bibitem [{\citenamefont {Cincio}\ and\ \citenamefont
  {Vidal}(2013)}]{Cincio-2012}%
  \BibitemOpen
  \bibfield  {author} {\bibinfo {author} {\bibfnamefont {L.}~\bibnamefont
  {Cincio}}\ and\ \bibinfo {author} {\bibfnamefont {G.}~\bibnamefont {Vidal}},\
  }\href {\doibase 10.1103/PhysRevLett.110.067208} {\bibfield  {journal}
  {\bibinfo  {journal} {Phys. Rev. Lett.}\ }\textbf {\bibinfo {volume} {110}},\
  \bibinfo {pages} {067208} (\bibinfo {year} {2013})}\BibitemShut {NoStop}%
\bibitem [{\citenamefont {Rezayi}\ and\ \citenamefont
  {Haldane}(1994)}]{Rezayi-1994}%
  \BibitemOpen
  \bibfield  {author} {\bibinfo {author} {\bibfnamefont {E.~H.}\ \bibnamefont
  {Rezayi}}\ and\ \bibinfo {author} {\bibfnamefont {F.~D.~M.}\ \bibnamefont
  {Haldane}},\ }\href {\doibase 10.1103/PhysRevB.50.17199} {\bibfield
  {journal} {\bibinfo  {journal} {Phys. Rev. B}\ }\textbf {\bibinfo {volume}
  {50}},\ \bibinfo {pages} {17199} (\bibinfo {year} {1994})}\BibitemShut
  {NoStop}%
\bibitem [{\citenamefont {Bergholtz}\ and\ \citenamefont
  {Karlhede}(2005)}]{Bergholtz-2005}%
  \BibitemOpen
  \bibfield  {author} {\bibinfo {author} {\bibfnamefont {E.~J.}\ \bibnamefont
  {Bergholtz}}\ and\ \bibinfo {author} {\bibfnamefont {A.}~\bibnamefont
  {Karlhede}},\ }\href {\doibase 10.1103/PhysRevLett.94.026802} {\bibfield
  {journal} {\bibinfo  {journal} {Phys. Rev. Lett.}\ }\textbf {\bibinfo
  {volume} {94}},\ \bibinfo {pages} {026802} (\bibinfo {year}
  {2005})}\BibitemShut {NoStop}%
\bibitem [{\citenamefont {Seidel}\ \emph {et~al.}(2005)\citenamefont {Seidel},
  \citenamefont {Fu}, \citenamefont {Lee}, \citenamefont {Leinaas},\ and\
  \citenamefont {Moore}}]{Seidel-2005}%
  \BibitemOpen
  \bibfield  {author} {\bibinfo {author} {\bibfnamefont {A.}~\bibnamefont
  {Seidel}}, \bibinfo {author} {\bibfnamefont {H.}~\bibnamefont {Fu}}, \bibinfo
  {author} {\bibfnamefont {D.-H.}\ \bibnamefont {Lee}}, \bibinfo {author}
  {\bibfnamefont {J.~M.}\ \bibnamefont {Leinaas}}, \ and\ \bibinfo {author}
  {\bibfnamefont {J.}~\bibnamefont {Moore}},\ }\href {\doibase
  10.1103/PhysRevLett.95.266405} {\bibfield  {journal} {\bibinfo  {journal}
  {Phys. Rev. Lett.}\ }\textbf {\bibinfo {volume} {95}},\ \bibinfo {pages}
  {266405} (\bibinfo {year} {2005})}\BibitemShut {NoStop}%
\bibitem [{\citenamefont {McCulloch}(2008)}]{McCulloch-2008}%
  \BibitemOpen
  \bibfield  {author} {\bibinfo {author} {\bibfnamefont {I.~P.}\ \bibnamefont
  {McCulloch}},\ }\href@noop {} {\enquote {\bibinfo {title} {Infinite size
  density matrix renormalization group, revisited},}\ } (\bibinfo {year}
  {2008}),\ \bibinfo {note} {unpublished},\ \Eprint
  {http://arxiv.org/abs/0804.2509} {arXiv:0804.2509} \BibitemShut {NoStop}%
\bibitem [{\citenamefont {Hastings}(2007)}]{Hastings-2007}%
  \BibitemOpen
  \bibfield  {author} {\bibinfo {author} {\bibfnamefont {M.~B.}\ \bibnamefont
  {Hastings}},\ }\href {http://stacks.iop.org/1742-5468/2007/P08024} {\bibfield
   {journal} {\bibinfo  {journal} {J. Stat. Mech.}\ }\textbf {\bibinfo {volume}
  {2007}},\ \bibinfo {pages} {P08024} (\bibinfo {year} {2007})}\BibitemShut
  {NoStop}%
\bibitem [{\citenamefont {Gottesman}\ and\ \citenamefont
  {Hastings}(2010)}]{Gottesman-2009}%
  \BibitemOpen
  \bibfield  {author} {\bibinfo {author} {\bibfnamefont {D.}~\bibnamefont
  {Gottesman}}\ and\ \bibinfo {author} {\bibfnamefont {M.~B.}\ \bibnamefont
  {Hastings}},\ }\href {\doibase 10.1088/1367-2630/12/2/025002} {\bibfield
  {journal} {\bibinfo  {journal} {New J. Phys.}\ }\textbf {\bibinfo {volume}
  {12}},\ \bibinfo {pages} {025002} (\bibinfo {year} {2010})}\BibitemShut
  {NoStop}%
\bibitem [{\citenamefont {Schuch}\ \emph {et~al.}(2008)\citenamefont {Schuch},
  \citenamefont {Wolf}, \citenamefont {Verstraete},\ and\ \citenamefont
  {Cirac}}]{Schuch-2008}%
  \BibitemOpen
  \bibfield  {author} {\bibinfo {author} {\bibfnamefont {N.}~\bibnamefont
  {Schuch}}, \bibinfo {author} {\bibfnamefont {M.~M.}\ \bibnamefont {Wolf}},
  \bibinfo {author} {\bibfnamefont {F.}~\bibnamefont {Verstraete}}, \ and\
  \bibinfo {author} {\bibfnamefont {J.~I.}\ \bibnamefont {Cirac}},\ }\href
  {\doibase 10.1103/PhysRevLett.100.030504} {\bibfield  {journal} {\bibinfo
  {journal} {Phys. Rev. Lett.}\ }\textbf {\bibinfo {volume} {100}},\ \bibinfo
  {eid} {030504} (\bibinfo {year} {2008})}\BibitemShut {NoStop}%
\bibitem [{\citenamefont {Nakamura}\ \emph {et~al.}(2012)\citenamefont
  {Nakamura}, \citenamefont {Wang},\ and\ \citenamefont
  {Bergholtz}}]{Nakamura-2012}%
  \BibitemOpen
  \bibfield  {author} {\bibinfo {author} {\bibfnamefont {M.}~\bibnamefont
  {Nakamura}}, \bibinfo {author} {\bibfnamefont {Z.-Y.}\ \bibnamefont {Wang}},
  \ and\ \bibinfo {author} {\bibfnamefont {E.~J.}\ \bibnamefont {Bergholtz}},\
  }\href {\doibase 10.1103/PhysRevLett.109.016401} {\bibfield  {journal}
  {\bibinfo  {journal} {Phys. Rev. Lett.}\ }\textbf {\bibinfo {volume} {109}},\
  \bibinfo {pages} {016401} (\bibinfo {year} {2012})}\BibitemShut {NoStop}%
\bibitem [{Note1()}]{Note1}%
  \BibitemOpen
  \bibinfo {note} {In the Supplemental materials we explain the numerical
  issues particular to quantum Hall DMRG, including the MPO construction and
  DMRG ergodicity issues. We also detail the computation of the quasiparticle
  charges, flux matrices, and modular $\protect \mathcal {T}$-matrix from the
  entanglement spectrum.}\BibitemShut {Stop}%
\bibitem [{\citenamefont {Schollw{\"o}ck}(2011)}]{Schollwock2011}%
  \BibitemOpen
  \bibfield  {author} {\bibinfo {author} {\bibfnamefont {U.}~\bibnamefont
  {Schollw{\"o}ck}},\ }\href {\doibase 10.1016/j.aop.2010.09.012} {\bibfield
  {journal} {\bibinfo  {journal} {Annals of Physics}\ }\textbf {\bibinfo
  {volume} {326}},\ \bibinfo {pages} {96} (\bibinfo {year} {2011})}\BibitemShut
  {NoStop}%
\bibitem [{\citenamefont {Moore}\ and\ \citenamefont
  {Read}(1991)}]{MooreRead1991}%
  \BibitemOpen
  \bibfield  {author} {\bibinfo {author} {\bibfnamefont {G.}~\bibnamefont
  {Moore}}\ and\ \bibinfo {author} {\bibfnamefont {N.}~\bibnamefont {Read}},\
  }\href {\doibase 10.1016/0550-3213(91)90407-O} {\bibfield  {journal}
  {\bibinfo  {journal} {Nuclear Physics B}\ }\textbf {\bibinfo {volume}
  {360}},\ \bibinfo {pages} {362 } (\bibinfo {year} {1991})}\BibitemShut
  {NoStop}%
\bibitem [{Note2()}]{Note2}%
  \BibitemOpen
  \bibinfo {note} {We note that for bosonic states the $\protect \mathds {1}$
  sector will not appear in the OES, a subtlety we address in the Supplementary
  materials. For the fermionic states studied here, $\protect \mathds {1}$
  will.}\BibitemShut {Stop}%
\bibitem [{\citenamefont {Bernevig}\ and\ \citenamefont
  {Haldane}(2008)}]{BernevigHaldane2008}%
  \BibitemOpen
  \bibfield  {author} {\bibinfo {author} {\bibfnamefont {B.~A.}\ \bibnamefont
  {Bernevig}}\ and\ \bibinfo {author} {\bibfnamefont {F.~D.~M.}\ \bibnamefont
  {Haldane}},\ }\href {\doibase 10.1103/PhysRevLett.100.246802} {\bibfield
  {journal} {\bibinfo  {journal} {Phys. Rev. Lett.}\ }\textbf {\bibinfo
  {volume} {100}},\ \bibinfo {pages} {246802} (\bibinfo {year}
  {2008})}\BibitemShut {NoStop}%
\bibitem [{\citenamefont {Wen}\ and\ \citenamefont
  {Wang}(2008)}]{WenWang-2008}%
  \BibitemOpen
  \bibfield  {author} {\bibinfo {author} {\bibfnamefont {X.-G.}\ \bibnamefont
  {Wen}}\ and\ \bibinfo {author} {\bibfnamefont {Z.}~\bibnamefont {Wang}},\
  }\href {\doibase 10.1103/PhysRevB.78.155109} {\bibfield  {journal} {\bibinfo
  {journal} {Phys. Rev. B}\ }\textbf {\bibinfo {volume} {78}},\ \bibinfo
  {pages} {155109} (\bibinfo {year} {2008})}\BibitemShut {NoStop}%
\bibitem [{\citenamefont {Rezayi}\ and\ \citenamefont
  {Haldane}(2000)}]{HaldaneRezayi2000}%
  \BibitemOpen
  \bibfield  {author} {\bibinfo {author} {\bibfnamefont {E.~H.}\ \bibnamefont
  {Rezayi}}\ and\ \bibinfo {author} {\bibfnamefont {F.~D.~M.}\ \bibnamefont
  {Haldane}},\ }\href {\doibase 10.1103/PhysRevLett.84.4685} {\bibfield
  {journal} {\bibinfo  {journal} {Phys. Rev. Lett.}\ }\textbf {\bibinfo
  {volume} {84}},\ \bibinfo {pages} {4685} (\bibinfo {year}
  {2000})}\BibitemShut {NoStop}%
\bibitem [{\citenamefont {Keski-Vakkuri}\ and\ \citenamefont
  {Wen}(1993)}]{Keski-Vakkuri-1993}%
  \BibitemOpen
  \bibfield  {author} {\bibinfo {author} {\bibfnamefont {E.}~\bibnamefont
  {Keski-Vakkuri}}\ and\ \bibinfo {author} {\bibfnamefont {X.-G.}\ \bibnamefont
  {Wen}},\ }\href {\doibase 10.1142/S0217979293003644} {\bibfield  {journal}
  {\bibinfo  {journal} {International Journal of Modern Physics B}\ }\textbf
  {\bibinfo {volume} {07}},\ \bibinfo {pages} {4227} (\bibinfo {year}
  {1993})}\BibitemShut {NoStop}%
\bibitem [{\citenamefont {Wen}\ and\ \citenamefont {Zee}(1992)}]{WenZee-1992}%
  \BibitemOpen
  \bibfield  {author} {\bibinfo {author} {\bibfnamefont {X.~G.}\ \bibnamefont
  {Wen}}\ and\ \bibinfo {author} {\bibfnamefont {A.}~\bibnamefont {Zee}},\
  }\href {\doibase 10.1103/PhysRevLett.69.953} {\bibfield  {journal} {\bibinfo
  {journal} {Phys. Rev. Lett.}\ }\textbf {\bibinfo {volume} {69}},\ \bibinfo
  {pages} {953} (\bibinfo {year} {1992})}\BibitemShut {NoStop}%
\bibitem [{\citenamefont {Levin}\ \emph {et~al.}(2007)\citenamefont {Levin},
  \citenamefont {Halperin},\ and\ \citenamefont {Rosenow}}]{Levin-aPf}%
  \BibitemOpen
  \bibfield  {author} {\bibinfo {author} {\bibfnamefont {M.}~\bibnamefont
  {Levin}}, \bibinfo {author} {\bibfnamefont {B.~I.}\ \bibnamefont {Halperin}},
  \ and\ \bibinfo {author} {\bibfnamefont {B.}~\bibnamefont {Rosenow}},\ }\href
  {\doibase 10.1103/PhysRevLett.99.236806} {\bibfield  {journal} {\bibinfo
  {journal} {Phys. Rev. Lett.}\ }\textbf {\bibinfo {volume} {99}},\ \bibinfo
  {pages} {236806} (\bibinfo {year} {2007})}\BibitemShut {NoStop}%
\bibitem [{\citenamefont {Lee}\ \emph {et~al.}(2007)\citenamefont {Lee},
  \citenamefont {Ryu}, \citenamefont {Nayak},\ and\ \citenamefont
  {Fisher}}]{Lee-aPf}%
  \BibitemOpen
  \bibfield  {author} {\bibinfo {author} {\bibfnamefont {S.-S.}\ \bibnamefont
  {Lee}}, \bibinfo {author} {\bibfnamefont {S.}~\bibnamefont {Ryu}}, \bibinfo
  {author} {\bibfnamefont {C.}~\bibnamefont {Nayak}}, \ and\ \bibinfo {author}
  {\bibfnamefont {M.~P.~A.}\ \bibnamefont {Fisher}},\ }\href {\doibase
  10.1103/PhysRevLett.99.236807} {\bibfield  {journal} {\bibinfo  {journal}
  {Phys. Rev. Lett.}\ }\textbf {\bibinfo {volume} {99}},\ \bibinfo {pages}
  {236807} (\bibinfo {year} {2007})}\BibitemShut {NoStop}%
\bibitem [{\citenamefont {Cincio}\ and\ \citenamefont {Vidal}()}]{CincioVidal}%
  \BibitemOpen
  \bibfield  {author} {\bibinfo {author} {\bibfnamefont {L.}~\bibnamefont
  {Cincio}}\ and\ \bibinfo {author} {\bibfnamefont {G.}~\bibnamefont {Vidal}},\
  }\href@noop {} {}\bibinfo {note} {In preparation}\BibitemShut {NoStop}%
\bibitem [{\citenamefont {{Tu}}\ \emph {et~al.}(2012)\citenamefont {{Tu}},
  \citenamefont {{Zhang}},\ and\ \citenamefont {{Qi}}}]{TuZhangQi}%
  \BibitemOpen
  \bibfield  {author} {\bibinfo {author} {\bibfnamefont {H.-H.}\ \bibnamefont
  {{Tu}}}, \bibinfo {author} {\bibfnamefont {Y.}~\bibnamefont {{Zhang}}}, \
  and\ \bibinfo {author} {\bibfnamefont {X.-L.}\ \bibnamefont {{Qi}}},\
  }\href@noop {} {\enquote {\bibinfo {title} {{Momentum polarization: an
  entanglement measure of topological spin and chiral central charge}},}\ }
  (\bibinfo {year} {2012}),\ \bibinfo {note} {unpublished},\ \Eprint
  {http://arxiv.org/abs/1212.6951} {arXiv:1212.6951 [cond-mat.str-el]}
  \BibitemShut {NoStop}%
\bibitem [{\citenamefont {Singh}\ \emph {et~al.}(2011)\citenamefont {Singh},
  \citenamefont {Pfeifer},\ and\ \citenamefont {Vidal}}]{SinghVidal-2011}%
  \BibitemOpen
  \bibfield  {author} {\bibinfo {author} {\bibfnamefont {S.}~\bibnamefont
  {Singh}}, \bibinfo {author} {\bibfnamefont {R.~N.~C.}\ \bibnamefont
  {Pfeifer}}, \ and\ \bibinfo {author} {\bibfnamefont {G.}~\bibnamefont
  {Vidal}},\ }\href {\doibase 10.1103/PhysRevB.83.115125} {\bibfield  {journal}
  {\bibinfo  {journal} {Phys. Rev. B}\ }\textbf {\bibinfo {volume} {83}},\
  \bibinfo {pages} {115125} (\bibinfo {year} {2011})}\BibitemShut {NoStop}%
\bibitem [{\citenamefont {{Kj{\"a}ll}}\ \emph {et~al.}(2012)\citenamefont
  {{Kj{\"a}ll}}, \citenamefont {{Zaletel}}, \citenamefont {{Mong}},
  \citenamefont {{Bardarson}},\ and\ \citenamefont {{Pollmann}}}]{Kjall-2013}%
  \BibitemOpen
  \bibfield  {author} {\bibinfo {author} {\bibfnamefont {J.~A.}\ \bibnamefont
  {{Kj{\"a}ll}}}, \bibinfo {author} {\bibfnamefont {M.~P.}\ \bibnamefont
  {{Zaletel}}}, \bibinfo {author} {\bibfnamefont {R.~S.~K.}\ \bibnamefont
  {{Mong}}}, \bibinfo {author} {\bibfnamefont {J.~H.}\ \bibnamefont
  {{Bardarson}}}, \ and\ \bibinfo {author} {\bibfnamefont {F.}~\bibnamefont
  {{Pollmann}}},\ }\href@noop {} {\enquote {\bibinfo {title} {{The phase
  diagram of the anisotropic spin-2 XXZ model: an infinite system DMRG
  study}},}\ } (\bibinfo {year} {2012}),\ \bibinfo {note} {unpublished},\
  \Eprint {http://arxiv.org/abs/1212.6255} {arXiv:1212.6255 [cond-mat.str-el]}
  \BibitemShut {NoStop}%
\bibitem [{\citenamefont {Verstraete}\ \emph {et~al.}(2004)\citenamefont
  {Verstraete}, \citenamefont {Garcia-Ripoll},\ and\ \citenamefont
  {Cirac}}]{Verstraete-2004}%
  \BibitemOpen
  \bibfield  {author} {\bibinfo {author} {\bibfnamefont {F.}~\bibnamefont
  {Verstraete}}, \bibinfo {author} {\bibfnamefont {J.~J.}\ \bibnamefont
  {Garcia-Ripoll}}, \ and\ \bibinfo {author} {\bibfnamefont {J.~I.}\
  \bibnamefont {Cirac}},\ }\href {\doibase 10.1103/PhysRevLett.93.207204}
  {\bibfield  {journal} {\bibinfo  {journal} {Phys. Rev. Lett.}\ }\textbf
  {\bibinfo {volume} {93}},\ \bibinfo {pages} {207204} (\bibinfo {year}
  {2004})}\BibitemShut {NoStop}%
\bibitem [{\citenamefont {McCulloch}(2007)}]{McCulloch-2007}%
  \BibitemOpen
  \bibfield  {author} {\bibinfo {author} {\bibfnamefont {I.~P.}\ \bibnamefont
  {McCulloch}},\ }\href {http://stacks.iop.org/1742-5468/2007/i=10/a=P10014}
  {\bibfield  {journal} {\bibinfo  {journal} {Journal of Statistical Mechanics:
  Theory and Experiment}\ }\textbf {\bibinfo {volume} {2007}},\ \bibinfo
  {pages} {P10014} (\bibinfo {year} {2007})}\BibitemShut {NoStop}%
\bibitem [{\citenamefont {Crosswhite}\ and\ \citenamefont
  {Bacon}(2008)}]{CrosswhiteBacon2008}%
  \BibitemOpen
  \bibfield  {author} {\bibinfo {author} {\bibfnamefont {G.~M.}\ \bibnamefont
  {Crosswhite}}\ and\ \bibinfo {author} {\bibfnamefont {D.}~\bibnamefont
  {Bacon}},\ }\href {\doibase 10.1103/PhysRevA.78.012356} {\bibfield  {journal}
  {\bibinfo  {journal} {Phys. Rev. A}\ }\textbf {\bibinfo {volume} {78}},\
  \bibinfo {pages} {012356} (\bibinfo {year} {2008})}\BibitemShut {NoStop}%
\bibitem [{\citenamefont {White}(2005)}]{White2005}%
  \BibitemOpen
  \bibfield  {author} {\bibinfo {author} {\bibfnamefont {S.~R.}\ \bibnamefont
  {White}},\ }\href {\doibase 10.1103/PhysRevB.72.180403} {\bibfield  {journal}
  {\bibinfo  {journal} {Phys. Rev. B}\ }\textbf {\bibinfo {volume} {72}},\
  \bibinfo {pages} {180403} (\bibinfo {year} {2005})}\BibitemShut {NoStop}%
\bibitem [{\citenamefont {Read}\ and\ \citenamefont
  {Rezayi}(1999)}]{ReadRezayi99}%
  \BibitemOpen
  \bibfield  {author} {\bibinfo {author} {\bibfnamefont {N.}~\bibnamefont
  {Read}}\ and\ \bibinfo {author} {\bibfnamefont {E.}~\bibnamefont {Rezayi}},\
  }\href {\doibase 10.1103/PhysRevB.59.8084} {\bibfield  {journal} {\bibinfo
  {journal} {Phys. Rev. B}\ }\textbf {\bibinfo {volume} {59}},\ \bibinfo
  {pages} {8084} (\bibinfo {year} {1999})}\BibitemShut {NoStop}%
\bibitem [{\citenamefont {Ginsparg}(1988)}]{Ginsparg}%
  \BibitemOpen
  \bibfield  {author} {\bibinfo {author} {\bibfnamefont {P.~H.}\ \bibnamefont
  {Ginsparg}},\ }\href@noop {} {\enquote {\bibinfo {title} {{Applied Conformal
  Field Theory}},}\ } (\bibinfo {year} {1988}),\ \Eprint
  {http://arxiv.org/abs/hep-th/9108028} {arXiv:hep-th/9108028} \BibitemShut
  {NoStop}%
\end{thebibliography}%
